\documentstyle[aaspp4,12pt]{article}

\newcommand{\etal}{{\em et~al. }}
\newcommand{\Msun}{{M$_\odot$ }}
\newcommand{\Msunns}{{M$_\odot$}}
\newcommand{\vsini}{{$v\sin i$ }}

\newcommand{\wcrit}{{$\omega_{crit}$}}
\newcommand{\wcrits}{{$\omega_{crit}$ }}

\begin{document}

\title{The Angular Momentum Evolution of Very Low Mass Stars}
\author{Alison Sills, M. H. Pinsonneault, D. M. Terndrup}
\affil{Department of Astronomy, The Ohio State University, 174 W. 18th
Ave., Columbus, OH, 43210, USA}

\begin{abstract}
We present theoretical models of the angular momentum evolution of
very low mass stars (0.1 - 0.5 \Msunns).  We also present models of
solar analogues (0.6 - 1.1 \Msunns) for comparison with previous work.
We investigate the effect of rotation on the effective temperature and
luminosity of these stars.  Rotation lowers the effective temperature and
luminosity of the models relative to standard models of the same mass
and composition. We find that the decrease in T$_{eff}$ and L can be
significant at the higher end of our mass range, but becomes small
below 0.4 \Msunns. The effects of different assumptions about internal
angular momentum transport are discussed. Formulae for relating
T$_{eff}$ to mass and v$_{rot}$ are presented. We demonstrate that the
kinetic energy of rotation is not a significant contribution to the
luminosity of low mass stars.

Previous studies of the angular momentum evolution of low mass stars
concentrated on solar analogues and were complicated by uncertainties
related to the internal transport of angular momentum.  In this paper
we extend our theoretical models for the angular momentum evolution of
stars down to 0.1 \Msunns.  We compare our models to rotational data
from young open clusters of different ages to infer the rotational
history of low mass stars, and the dependence of initial conditions
and rotational evolution on mass. We find that the qualitative
conclusions for stars below 0.6 \Msun do not depend on the assumptions
about internal angular momentum transport with the exception of a
zero-point shift in the angular momentum loss saturation threshold.
We argue that this makes these low mass stars ideal candidates for the
study of the angular momentum loss law and distribution of initial
conditions.  For stars with masses between 0.6 and 1.1 \Msun, scaling
the saturation threshold by the Rossby number can reproduce the
observed mass dependence of the stellar angular momentum evolution.
We find that neither models with solid body rotation nor
differentially rotating models can simultaneously reproduce the
observed stellar spin down in the 0.6 to 1.1 \Msun mass range and for
stars between 0.1 and 0.6 \Msunns.  We argue that the most likely
explanation is that the saturation threshold drops more steeply at low
masses than would be predicted with a simple Rossby scaling.  In young
clusters there is a systematic increase in the mean rotation rate with
decreased temperature below 3500 K (0.4 \Msunns).  This suggests
either inefficient angular momentum loss or mass-dependent initial
conditions for stars near the fully convective boundary.
\end{abstract}

\keywords{stars: evolution -- stars: rotation -- stars: interiors --
stars: formation -- low mass stars}

\section{Introduction}

The study of stellar rotation can provide insights into a variety of
interesting subjects in the fields of star formation and stellar
structure and evolution.  The observed rotation velocities and
rotation periods of open cluster stars as a function of mass and age
yield clues about the star formation process, the internal transport
of angular momentum, the loss of angular momentum through magnetized
stellar winds, and the origin and generation of stellar magnetic
fields.  In addition, rotation can drive mixing not present in
standard stellar models with important consequences for the observed
surface abundances of stars.

We are now able to observe significant samples of stars down to the
hydrogen burning limit in open clusters (e.g. NGC 2420, von Hippel
\etal 1996), globular clusters (e.g. 47 Tucanae, Santiago, Elson \&
Gilmore 1996), and the field (e.g. Tinney, Mould \& Reid 1993). We
have also been able to observe rotation in these stars, using both
spectroscopy to determine \vsini (Kraft 1965, Stauffer \etal 1997a),
and photometry to monitor spot modulation on the stars
(Barnes \etal 1999, Prosser \etal 1995) and thereby determine
rotational periods. This plethora of information about the rotation of
low mass stars has been a great boon to the study of these stars for a
number of reasons. First of all, the rotation rates of stars on the
main sequence are determined by their pre-main sequence evolution, so
that by studying the rotational evolution of low mass stars, we can
investigate the early stages of stellar evolution. Second, stellar
magnetic phenomena are related to stellar rotation. The evolution of the
rotation rates is largely determined by angular momentum loss from a
magnetized stellar wind (Kawaler 1988, Weber \& Davis 1967). Stellar
rotation is found to correlate with chromospheric activity and other
magnetic tracers (for a review see Hartmann \& Noyes 1987), which
lends support to the idea that rotation plays a crucial role in the
generation of stellar magnetic fields, through the operation of a dynamo.

In recent years there have been a large number of observational and
theoretical studies of the angular momentum evolution of low mass
stars (see Krishnamurthi \etal 1997 for a review).  These studies
have focused on solar analogues which have the most extensive
observational database.  However, the majority of the moment of
inertia of solar analogues is in the radiative interior, so theoretical
predictions for their angular momentum evolution are strongly influenced
by the treatment of internal angular momentum transport.  In this
paper we examine the rotational properties of stellar models of lower
mass stars where the radiative core provides a smaller (or even
nonexistent) fraction of the moment of inertia.  We will show that the
study of these low mass stars can provide valuable clues about angular
momentum loss and the distribution of initial conditions that depend
only weakly on the treatment of internal angular momentum transport.
We begin with a discussion of the important ingredients for
theoretical models of stellar angular momentum evolution.

The rotational evolution of a low mass star is determined by four
factors. The first factor is the star's initial angular momentum.
Young low mass stars are fully convective, and helioseismic data
indicate that the rotation rate in the solar convective zone is
independent of radius; as a result, the initial angular momentum is
fully specified by the initial rotation period.  Observed T Tauri star
rotation periods are between 2 and 16 days, with an average of 9.54
days and a median of 8.5 days (Choi \& Herbst 1996).

The rate at which angular momentum is lost through a magnetic wind
will also strongly influence the stellar rotation rate.  In a linear
dynamo the angular momentum loss rate is proportional to the surface
angular velocity cubed (Weber \& Davis 1967, Kawaler 1988).  However,
an angular momentum loss law of this form predicts rapid spin down of
fast rotators in disagreement with observational data (Pinsonneault et
al. 1990).  More recent theoretical models use an angular momentum
loss law that saturates at some critical rotation rate; the lower the
saturation threshold, the longer that rapid rotation can
persist. Previous work on solar analogues (Barnes \& Sofia 1996,
Krishnamurthi \etal 1997) has suggested that the saturation threshold
depends on mass, and is inversely proportional to the global
convective overturn timescale. This scaling is prompted by its
relationship to the Rossby number, a measure of the strength of dynamo
magnetic activity in the star, and supported by observational data on
the relationship between chromospheric and coronal activity
indicators, rotation, and mass (e.g. Noyes \etal 1984, Patten \& Simon
1996, Krishnamurthi \etal 1998).

The rotation history of a star is also influenced by the internal
transport of angular momentum.  If a torque is applied to the surface
convection zone a shear will be generated between the convective
envelope and radiative core; the angular momentum evolution will
depend on the efficiency of the angular momentum coupling between the
core and envelope.

One straightforward model is to assume that the coupling time scale is
extremely short, i.e. that solid body rotation is enforced at all
times throughout the star.  This is predicted in the limiting case of
strong magnetic coupling between the core and envelope (see for
example Charbonneau \& MacGregor 1993), and models of this type have
been investigated in the context of the angular momentum evolution of
solar analogues (Collier-Cameron \& Jianke 1994; Bouvier, Forestini,
\& Allain 1997; Krishnamurthi et al. 1997).

Another approach is to solve for the internal transport of angular
momentum by hydrodynamic mechanisms (see Pinsonneault 1997 for a
review), internal gravity waves (Talon \& Zahn 1998), magnetic fields
(Charbonneau \& MacGregor 1993), or hybrid models including more than
one of these mechanisms (e.g. Barnes, Charbonneau, \& MacGregor 1999).
In these cases, angular momentum can be stored in a rapidly
rotating core.  The rotational evolution of models of this class can
therefore be different than that of solid body models as that
reservoir of angular momentum can be shielded from the strong angular
momentum loss that accompanies rapid surface rotation.

One of the most challenging features of the open cluster data for
theoretical models to explain is the large number of young slowly
rotating stars.  A simple projection of T Tauri star rotation
velocities with conservation of angular momentum would predict more
rapid rotation than is observed, and the short time scale would not
permit sufficient angular momentum loss given the modest predicted
rotation rates (see for example Stauffer \& Hartmann 1987, Keppens,
MacGregor, \& Charbonneau 1995).  K\"{o}nigl (1991) proposed that the
presence of an accretion disk, coupled to a protostar by a magnetic
field, will force the protostar to rotate with the same period as the
disk (see also Keppens, MacGregor, \& Charbonneau 1995;
Collier-Cameron, Campbell, \& Quaintrell 1995). Only when the disk is
disrupted can the star begin its normal angular momentum evolution
(spin-up because of contraction modified by the processes described
above).  When the star is no longer locked to the disk, it begins to
spin up as its radius contracts. Since the star is starting with a
smaller moment of inertia, it can never spin up as much as a star
which lost its disk at the birthline. This would produce a range of
surface rotation rates from a range of accretion disk lifetimes and
provided an attractive physical explanation for the observed
distribution of stellar surface rotation rates.  There is indirect
evidence for the model (Edwards et al. 1993, Choi \& Herbst 1996), in
the sense that the distribution of rotation periods is different for
young stars with infrared excesses than for young stars without them
(but see also Stassun \etal 1999).  The longest reasonable disk
lifetime, as determined from studies of T Tauri stars, is about 10 Myr
(Strom \etal 1989).  We therefore wish to reproduce the slowest
rotators in each cluster with models which have disk lifetimes of 10
Myr or less.  This creates problems for solid body models, since they
require much longer disk lifetimes and also have difficulty in
explaining the observed spindown of slow rotators on the main sequence
(Krishnamurthi \etal 1997, Allain 1998).

The ingredients which determine the rotational evolution of low mass
stars are not independent. For example, a rapidly rotating star could
exist because it had a short initial period. It could be rotating
rapidly because it has not lost much angular momentum since the wind
saturation threshold is low. Or, it could have been locked to its disk
for a very short period of time. However, this degeneracy of factors
is not an insoluble problem.  We can determine the angular momentum
loss rate (the saturation threshold) by comparing rotation rates of
the fastest rotators in clusters of different ages. These stars must
have had the fastest initial periods and have not been locked to
disks. Therefore, any change in their rotation rates will be caused by
their contraction to the main sequence, plus any loss due to their
wind.  We have evidence from solar analogues that the magnetic
wind saturation threshold depends on mass (Barnes \& Sofia 1996,
Krishnamurthi \etal 1997). By modeling the fastest rotators in a
series of clusters with different ages, we can constrain this
threshold for each mass.  After we know the dependence of the
saturation threshold on mass, we can attempt to model the slowest
rotators in young open clusters by allowing them to retain their disks
for long periods of time.

By studying the lowest mass stars, we can simplify the problem.  Stars
with masses less than $\sim$ 0.5 \Msun are fully convective throughout
their pre-main sequence evolution, while stars with masses less than
0.25 \Msun are fully convective for their entire lifetime.
Helioseismic data suggest that they will
therefore always rotate as solid bodies, so the internal transport of
angular momentum in such stars can be modeled very simply.  We can use
fully convective stars to diagnose the initial conditions in open
clusters, such as the range of disk lifetimes and initial rotation
periods, and the mass dependence of the angular momentum loss law.

As discussed above, a number of researchers have investigated the
angular momentum evolution of solar analogues (0.8 - 1.2 \Msunns). The
major obstacles which prevented us from modeling very low mass stars
accurately in the past have been the lack of adequate model
atmospheres, opacities and equations of state for low temperatures
(less than 4000 K). Lately, several groups (Alexander \& Ferguson
1994; Allard \& Hauschildt 1995; Saumon, Chabrier \& Van Horn 1995)
have made breakthroughs in the necessary physics. Improved
evolutionary models of very low mass stars have been produced over the
last few years (Baraffe \etal 1998). However, the effects of rotation
have not been included in these recent low mass models, and neglecting
rotation could lead to anomalous results. For example, rotation can
modify the amount of lithium depletion in low mass stars, affecting
the derived ages from lithium isochrone fitting (e.g. Stauffer,
Schultz \& Kirkpatrick 1998). In this work, we present the first
rotational models of stars less massive than 0.5 \Msunns. We also
include models for stars up to 1.1 \Msunns, for comparison with
previous work. In section 2, we discuss the methods used to determine
the rotational evolution of the low mass stars. We present the results
in section 3, and discuss their implications in section 4.

\section{Method}

We used the Yale Rotating Stellar Evolution Code (YREC) to construct
models of the low mass stars. YREC is a Henyey code which solves the
equations of stellar structure in one dimension (Guenther \etal
1992). The star is treated as a set of nested, rotationally deformed
shells. The chemical composition of each shell is updated separately
using the nuclear reaction rates of Gruzinov \& Bahcall (1998). The
initial chemical mixture is the solar mixture of Grevesse \& Noels
(1993), and our models have a metallicity of Z=0.0188. Gravitational
settling of helium and heavy elements is not included in these
models. We use the latest OPAL opacities (Iglesias \& Rogers 1996) for
the interior of the star down to temperatures of $\log T (K) = 4$. For
lower temperatures, we use the molecular opacities of Alexander \&
Ferguson (1994).  For regions of the star which are hotter than $\log
T (K) \geq 6$, we used the OPAL equation of state (Rogers, Swenson \&
Iglesias 1996). For regions where $\log T (K) \leq 5.5$, we used the
equation of state from Saumon, Chabrier \& Van Horn (1995), which
calculates particle densities for hydrogen and helium including
partial dissociation and ionization by both pressure and temperature.
In the transition region between these two temperatures, both
formulations are weighted with a ramp function and averaged.  The
equation of state includes both radiation pressure and electron
degeneracy pressure. For the surface boundary condition, we used the
stellar atmosphere models of Allard \& Hauschildt (1995), which
include molecular effects and are therefore relevant for low mass
stars.  We used the standard B\"{o}hm-Vitense mixing length theory
(Cox \& Guili 1968; B\"{o}hm-Vitense 1958) with $\alpha$=1.72.  This
value of $\alpha$, as well as the solar helium abundance,
$Y_{\odot}=0.273$, was obtained by calibrating models against
observations of the solar radius ($6.9598 \times 10^{10}$ cm) and luminosity
($3.8515 \times 10^{33}$ erg/s) at the present age of the Sun (4.57
Gyr).

The structural effects of rotation are treated using the scheme
derived by Kippenhahn \& Thomas (1970) and modified by Endal \& Sofia
(1976). The details of this particular implementation are discussed in
Pinsonneault \etal (1989). In summary, quantities are evaluated on
equipotential surfaces rather than the spherical surfaces usually used
in stellar models. The mass continuity equation is not altered by
rotation:
\begin{equation}
\frac{\partial M}{\partial r} = 4\pi r^2 \rho.
\end{equation}
The equation of hydrostatic equilibrium includes a term which takes
into account the modified gravitational potential of the non-spherical
equipotential surface:
\begin{equation}
\frac{\partial P}{\partial M} = - \frac{GM}{4 \pi r^4} f_P,
\end{equation}
where
\begin{equation}
f_P = \frac{4\pi r^4}{GMS} \frac{1}{\langle g^{-1}\rangle},
\end{equation}
and
\begin{equation}
\langle g^{-1} \rangle = \frac{1}{S} \int_{\psi = const} g^{-1} d\sigma,
\end{equation}
$S$ is the surface area of an equipotential surface, and $d\sigma$ is
an element of that surface. The factor $f_P$ is less than one for
non-zero rotation, and approaches one as the rotation rate goes to
zero.
The radiative temperature gradient also depends on rotation:
\begin{equation}
\frac {\partial \ln T}{\partial \ln P} = \frac{3 \kappa}{16 \pi acG}\frac{P}{T^4}\frac{L}{M}\frac{f_T}{f_P},
\end{equation}
where
\begin{equation}
f_T = \left(\frac{4\pi r^2}{S}\right)^2 \frac{1}{\langle g \rangle
\langle g^{-1} \rangle},
\end{equation}
and $\langle g \rangle$ is analogous to $\langle g^{-1}\rangle$. $f_T$
has the same asymptotic behaviour as $f_P$, but is typically much
closer to 1.0. The energy conservation equation retains its
non-rotating form. Therefore, all the structural effects of rotation
are limited to the equation of hydrostatic equilibrium and the
radiative temperature gradient. This modified temperature gradient is
used in the Schwarzschild criterion for convection:
\begin{equation}
\frac{\partial \ln T}{\partial \ln P} = min \left[\nabla_{ad},\nabla_{rad} \frac{f_T}{f_P} \right]
\end{equation}
where $\nabla_{ad}$ and $\nabla_{rad}$ are the normal spherical adiabatic
and radiative temperature gradients.

The Endal \& Sofia scheme is valid across a wide range of rotation
rates, for a restricted class of angular momentum distributions. This
scheme requires that the rotational velocity is constant on
equipotential surfaces, which does not allow for modeling of
latitude-dependent rotational profiles. We assume that horizontal
turbulence is sufficiently strong to enforce spherical rotation
(Chaboyer \& Zahn 1992). The other restriction on the Endal \& Sofia
scheme is that it assumes that the potential is conservative, which is
not valid when the star is expanding or contracting. Therefore, it is
necessary to take small timesteps during any phase of expansion or
contraction (such as the pre-main sequence) to minimize the errors
introduced by this limitation. This method, and most others used in
rotational stellar evolution codes, does not include the horizontal
transport of heat, which may be important in the most rapidly rotating
stars, rotating very close to their breakup velocities. See Meynet \&
Maeder (1997) for a detailed discussion of the validity of this
approach to the evolution of rotating stars.

To model the loss of angular momentum from the surface, we adopt a
modified Kawaler angular momentum loss rate with a N=1.5 wind law
(Chaboyer, Demarque \& Pinsonneault 1995), given by
\begin{equation}
\frac{dJ}{dt} =  -K \omega^3 \left(\frac{R}{R_{\odot}}\right)^{0.5}
\left(\frac{M}{M_{\odot}}\right)^{-0.5} , \; \omega \leq \omega_{crit} 
\end{equation}

\[ \frac{dJ}{dt} =  -K \omega_{crit}^2 \omega
\left(\frac{R}{R_{\odot}}\right)^{0.5}
\left(\frac{M}{M_{\odot}}\right)^{-0.5} , \; \omega > \omega_{crit} \]

This represents the draining of angular momentum from the outer
convection zone through a magnetic wind. The magnetic field is assumed
to be proportional to the angular velocity while that velocity is
small, but then saturates at $\omega = \omega_{crit}$.  The constant
$K$ is calibrated by reproducing the solar rotation rate (2 km/s) at
the solar age (4.57 Gyr from the birthline) for a 1.0 \Msun model with
an initial period of 10 days and no disk-locking. The value of
$\omega_{crit}$ has been found to depend on mass (Krishnamurthi \etal
1997, Barnes \& Sofia 1996). We have used the prescription for the
variation of $\omega_{crit}$ with mass from Krishnamurthi \etal
(1997). In this prescription, $\omega_{crit}$ is inversely
proportional to the convective overturn timescale in the star at 200
Myr (Kim \& Demarque 1996):
\begin{equation}
\omega_{crit} = \omega_{crit\odot} \frac{\tau_{\odot}}{\tau}
\end{equation}
The convective overturn times were linearly extrapolated for masses
lower than 0.5 \Msunns. We have also considered models in which no
angular momentum loss is allowed, to determine the maximum structural
effects of rotation.

We have calculated the rotational evolution of two classes of stellar
models: models in which solid body rotation is enforced at all time,
and models in which internal angular momentum transport is affected by
hydrodynamic mechanisms. The rotation rate of the solid body models is
determined from the moment of inertia of the model at a given time,
and the total angular momentum as determined by the loss rate given in
equation 8. For the second set of models, rigid rotation is enforced
throughout the convection zone only, and the rotation in the interior
is governed by the transport of angular momentum by hydrodynamic
mechanisms. The chemical mixing associated with this angular momentum
transport is computed using a set of diffusion equations (Pinsonneault
\etal 1989); the amount of coupling between transport and mixing is
calibrated by requiring that the amount of lithium depletion
calculated by our model matches the observed value for the Sun.

We start our models on the birthline of Palla \& Stahler (1991), which
is the deuterium-burning main sequence and corresponds to the upper
envelope of T Tauri observations in the HR diagram. It has been shown
(Barnes \& Sofia 1996) that this physically realistic assumption for
the initial conditions of stellar rotation models is crucial for
accurate modeling of ultra-fast rotators in young clusters. The models
with no angular momentum loss began with an initial rotation period of
8 days, corresponding to the mean classical T Tauri star rotation
period (Choi \& Herbst 1996). The models which included angular
momentum loss began with initial rotation periods of either 4 or 10
days. We present models for solar metallicity stars between 0.1 and
1.1 \Msun in increments of 0.1 \Msunns. These models have been evolved
from the birthline to an age of 10 Gyr.

\section{Results}

\subsection{The Structural Effects of Rotation}

Evolutionary tracks for both rotating and non-rotating models are
presented in figure 1. The rotating models have initial periods of 10
days and experience no angular momentum loss. As expected (Sackmann
1970, Pinsonneault \etal 1989), the effect of rotation is to shift
stars to lower effective temperatures and lower luminosities,
mimicking a star of lower mass. This effect is most pronounced for the
highest mass stars presented in this paper, and is reduced to a low
level for stars less than 0.4 \Msunns. Since low mass stars are fully
convective, their temperature gradient will be the adiabatic gradient,
which does not depend on the rotation rate (equation 7). However, the
structural effects of rotation are still apparently in fully
convective stars, and diminish with decreasing mass. This suggests
that an additional mechanism is also at work.  As stars get less
massive, their central pressure is being provided less by thermal
pressure and more by degeneracy pressure. The amount of degeneracy is
determined by the density in the interior, which is not affected by
rotation (see equation 1). Rotation provides an additional method of
support for the star, but in stars with a significant amount of
degeneracy, the rotational support is a smaller fraction of the total
pressure. Therefore, the structure of the low mass stars is less
affected by rotation than their higher mass counterparts.

Figure 2 compares the evolutionary tracks for rotating stars under
different assumptions about internal angular momentum transport. The
solid tracks are stars which have differentially rotating radiative
cores and rigidly rotating convection zones, while the dashed lines
show the tracks for stars which are constrained to rotate as solid
bodies. The two tracks for each mass have the same surface rotation
rate at the zero-age main sequence. The low mass stars show no
difference between the two assumptions, since these stars are fully
convective for the entire 10 Gyr plotted here. Therefore, they always
rotate as solid bodies. The higher mass stars begin their lives high
on the pre-main sequence as fully convective, solid body rotators. As
they contract and develop radiative cores, however, the difference in
the two assumptions about angular momentum transport becomes
apparent. Differential rotators have a higher total angular momentum
than solid body rotators of the same surface rotation rate. As stars
contract along the pre-main sequence, they become more centrally
concentrated, which means that the core spins up more than the
envelope does.  The solid body rotators are forced to spread their
angular momentum evenly throughout the star, so they have less total
angular momentum for a given surface rotation rate.  Therefore, the
impact of rotation on the structure of the star is larger for
differential rotators than for solid body rotators of the same surface
rotation rate.  However, at constant initial angular momentum, the
solid body rotators are cooler at the zero age main sequence, and have
longer pre-main sequence lifetimes, than differentially rotating stars
of the same mass.  When comparing the effects of rotation between
different models, it is important to note whether the comparison is
between stars with the same current surface rotation rate, or with the
same initial angular momentum.

We included the kinetic energy of rotation ($T=\frac{1}{2}I\omega^2$)
in our determination of the total luminosity in each shell of the
star. As the star changes its moment of inertia $I$ and its rotation
rate $\omega$, the resulting change in its rotational kinetic energy
can be included in the energy budget of the star.  Most
implementations of stellar rotation into stellar structure and
evolution neglect this energy since it is expected that the amount of
kinetic energy available is not enough to significantly affect the
evolution of the star. Since very low mass stars have much lower
luminosities than solar-mass stars, but their moments of inertia are
not as significantly lower, it is plausible that the kinetic energy of
rotation would contribute a significant fraction of the total
luminosity of the star.  As shown in figure 4, however, the change in
the kinetic energy of rotation contributes no more than 6\% of the
total luminosity of the star in the 1.0 \Msun model, and that
contribution lasts less than 50 Myr. As expected, the lowest mass star
has the most significant contribution, lasting for about 1 Gyr, but at
4\% or less. The positions of stars in the HR diagram are minimally
affected by the inclusion of this source of energy. The kinetic energy
of rotation reduces the luminosity at any given time by less than 0.02
dex in $\log(L/L_{\odot})$, and usually less than 0.005 dex. The
timescales for evolution are also equally unaffected. The models with
no angular momentum loss will have the maximum possible effect of
rotational kinetic energy. Since these models show no significant
effect, we conclude that the change in the kinetic energy of rotation
is at most a perturbation on the structure.

The main structural effect of rotation is a reduction in the effective
temperature of stars. Using our tracks, we have quantified the
relationship between rotational velocity and the difference in
effective temperature at the zero age main sequence. In figure 5 we
present this relationship for stars of different masses, and for both
the differentially rotating (solid lines) and solid body models
(dashed lines). For low mass stars, the difference in temperature
caused by rotation is of order a few tens of K (and reduces to
less than 10 K for stars of 0.2 \Msunns). This difference is
therefore negligible. However, the reduction in effective temperature
is larger for the more massive stars, and can reach significant levels
of a few hundred K for stars more massive than about 0.6
\Msunns. Therefore, when determining masses from observed temperatures
or colours, it is important know how fast these stars are
rotating. The relationship between rotation rate and difference in
effective temperature, for a given stellar mass, is well-fit by a
polynomial. The coefficients for this polynomial at different masses
and under different assumptions of internal angular momentum transport
are given in table 2. It should be noted that while solid body
rotators of the same initial period rotate faster at the zero age main
sequence than differentially rotating stars, the structural effects of
rotation are slightly more pronounced in the differential rotators at
constant surface rotation speed. Therefore, for a constant rotational
velocity, stars which rotate differentially have a higher angular
momentum than solid body rotators.

For stars of the same mass, rotation reduces the luminosity of stars
as well as their temperatures. The difference in luminosity is not as
important as the difference in temperature cause by rapid rotation, as
shown in figure 5. Even for the most extreme case, the difference in
luminosity for a 1.0 \Msun star rotating at 250 km/s is less than 0.12
dex in $\log(L_{\odot})$. While differences of this size will result
in a thicker main sequence of a cluster, it should not affect any
scientific results significantly. Most stars in clusters do not rotate
very fast, so the upper main sequence will be well-defined for any
isochrone fitting or distance determination. Luminosity is used as an
indicator of mass for low mass stars, but since the difference in
luminosity between rapid rotators and non-rotators is very small for
low mass stars, this calibration should not be affected by rotation.

The total effect of rotation is such that the locus of the zero age
main sequence becomes brighter as stars rotate more quickly. The
combination of a significant decrease in temperature with a small
decrease in luminosity for stars of the same mass moves the locus
above the non-rotating main sequence. At a surface rotation rate of
100 km/s, the rotating main sequence is brighter by about 0.01
magnitudes. At 200 km/s, the sequence is brighter by 0.03 magnitudes.
Therefore, we expect to see rapid rotators in clusters lying above the
cluster main sequences by a few hundredths of a magnitude.

Since they are fainter, rapid rotators have slightly longer lifetimes
compared to non-rotating stars of the same mass. The amount of
increase depends on the mass of the star and the rotation rate, but in
the most extreme case (1.0 \Msun rotating at 250 km/s), the difference
in pre-main sequence lifetime is 7\%. For rotation rates less than 100
km/s, the increase in lifetime is less than 1\% for all masses.

\subsection{Rotational Evolution}

We have compared our models with rotational data from young open
clusters.  Each cluster provides a sample of stars with different
masses, allowing us to study the effects of mass on angular momentum
loss and disk lifetimes. By comparing the progression of rotation
rates from very young clusters to older ones, we can study the
rotational history of stars of a range of masses. Both probes are very
useful in the study of stellar rotation.  There are four well-studied
data sets used when investigating rotation in young open clusters: IC
2602 and IC 2391, at 30 Myr; $\alpha$ Persei, at 60 Myr, the Pleiades
at 110 Myr; and the Hyades, at 600 Myr. In figures 6-16, we compare
our models to observational data at the appropriate age for the
cluster, under a number of different assumptions about the initial
conditions, the internal transport of angular momentum, the saturation
threshold, and the disk locking lifetime. The data were taken from
Stauffer \etal 1997b (IC 2602 and IC 2391), Prosser \etal 1995 and
references therein ($\alpha$ Per), Soderblom \etal 1993; Prosser \etal
1995 and Stauffer \etal 1999 (Pleiades and Hyades), and Radick \etal
1987 (Hyades), supplemented with data from the Open Cluster Database
(Prosser \& Stauffer 1999).  

We wish to reproduce a number of important features of these data
sets. First, our models must predict the correct rotation rates for the
fast rotators in each of these clusters. The presence of fast rotators
is caused by the saturation of the angular momentum loss law (equation
8), and constrains the value of $\omega_{crit}$ as a function of
mass. Also, our models should reproduce the spin-down of the fast
rotators with time, as seen by comparing stars of the same mass in
different clusters.

The comparison between our models and observed fast rotators confirms
the conclusion of Krishnamurthi \etal (1997) that a mass-dependent
$\omega_{crit}$ is necessary, in the sense that $\omega_{crit}$
increases for increasing mass. Figures 6 and 7 present rotational
models with different normalizations of the Rossby scaling and initial
periods of 10 days. Figure 6 presents models which rotate as solid
bodies throughout their lifetime, while figure 7 shows differentially
rotating models in which internal angular momentum transport is
determined by hydrodynamical instabilities.  The thick solid lines in
each frame represent the upper envelope of rotation rates observed in
each cluster, which have not been corrected for inclination
angle. Observations of X-ray activity in rotating stars as a function
of \vsini show that reasonable values for the angular momentum loss
saturation threshold velocity range from 5 to 20 $\omega_{\odot}$
(Patten \& Simon 1996). We have plotted three different
normalizations, with values of \wcrit$_{\odot}$ = 5, 10 and 20
$\omega_{\odot}$ (from top to bottom each frame). For the solid body
models (figure 6), it is clearest at the older ages that stars with a
high saturation threshold lose too much angular momentum, and the low
saturation threshold models lose too little. Therefore, the best
choice for normalization is about 10 $\omega_{\odot}$. For the
differentially rotating models, the best choice for normalization is
about 5 $\omega_{\odot}$, since the other values of \wcrit$_{\odot}$
produce stars which are rotating too slowly. These normalizations are
the same as the ones adopted by Krishnamurthi \etal (1997) for solar
analogues.

The Rossby scaling with the normalizations suggested by Krishnamurthi
\etal (1997) are shown in figures 8 (solid body models) and 9
(differentially rotating models). The upper line in each frame
corresponds to models with initial rotation periods of 4 days, while
the lower line corresponds to models with initial periods of 10
days. The fastest rotators in all the clusters lie below the 4 day
line, with the exception of the lowest mass stars in the Hyades, which
rotate faster than the predictions of the differentially rotating
models. We conclude that a different normalization for the Rossby
scaling is necessary for low mass stars.

The second main feature of these data sets that we wish to reproduce
is the large spread in rotation rate at constant mass. This range in
rotational velocity is caused by a range in protostellar disk
lifetime.  In figures 10 and 11, we present models with initial
rotation periods of 10 days, and the above-mentioned normalizations
for \wcrit. The upper line in each frame shows models which have
detached from their disk at the birthline. Moving down each frame, the
lines represent models with disk lifetimes of 0.3, 1, 3 and 10 Myr
from the birthline. The differentially rotating models (figure 11)
reproduce the rotation rates of the slowest rotators in these clusters
with disk lifetimes of 10 Myr or less, while the solid body models
(figure 10) require longer disk lifetimes, perhaps even as long as the
current age of the cluster.

We have verified that neither the Krishnamurthi \etal (1997) Rossby
scaling nor any other unique Rossby scaling can reproduce the mass
dependence of the angular momentum evolution below 4000 K,
corresponding to masses of 0.6 \Msun and below.  The persistence of
rapid rotation in the Hyades cluster requires inefficient angular
momentum loss in the lowest mass stars, while a uniform lowering of
the saturation threshold would predict more rapid rotation for higher
mass stars than is observed in the Hyades.  We therefore constructed
models for the 0.1 to 0.5 \Msun range where the value of \wcrit was
tuned to reproduce the Hyades upper envelope; we adopted the
Krishnamurthi \etal (1997) Rossby scaling for the higher mass models.

Figures 12 and 13 show the same models as in figures 10 and 11, but
with this different normalizations of the mass dependence for
\wcrit. In figure 12, we present solid body rotational models with
\wcrit$_{\odot}$ = 7, 6.2, 5.1, 3.6, 1.7 $\omega_{\odot}$ at 0.5, 0.4,
0.3, 0.2 and 0.1 \Msun respectively. Figure 13 shows differentially
rotating models with \wcrit$_{\odot}$ = 3, 3.3, 3.5, 1.9, 0.9
$\omega_{\odot}$ for the same masses. Both these sets of models
reproduce the fastest rotators in the Hyades at low masses. With the
exception of a different zero-point for the saturation threshold, the
solid body models and the differentially rotating models predict very
similar angular momentum evolution histories for these low mass stars.
The different normalization and similar evolution can be understood as
follows.  The DR models require a higher constant in the angular
momentum loss law to extract angular momentum from a rapidly rotating
core and a slowly rotating envelope.  For models with $\omega >$
\wcrit, the angular momentum loss rates can be made identical by a
suitable zero-point shift in \wcrit.  Although the DR models would
experience more severe angular momentum loss for $\omega < $\wcrit,
the low values of \wcrits required for reproducing the stellar
rotation data are not reached until ages older than the clusters that
we are studying.

In figure 14 we use the models presented in the previous two figures
to produce distributions of disk lifetimes for three of the clusters
studied in this paper. A statistical correction of 4/$\pi$ was applied
to the observed rotation velocities. This overestimates the rotation
rates for the rapid rotators, where $\sin i$ is likely to be close to
one, but does provide a more accurate estimate on average for the bulk
of the slow rotator population. We find an essentially constant
distribution of disk lifetimes with age for the differentially
rotating models across the young clusters.  Disk lifetimes longer than
10 Myr are inconsistent with observations of infrared excesses around
T Tauri stars (Strom \etal 1989), which predict maximum disk lifetimes
between 3 and 10 Myr. Therefore, the large fraction of stars which are
required to have long disk lifetimes is an argument against the solid
body models. 

The predicted rotation rates at the age of the Hyades are
systematically slightly higher than the data for the differentially
rotating models at the higher end of the mass range, while the solid
body models are in good agreement with the observed range.  This leads
to spuriously short disk lifetimes for the higher mass stars when
combined with the insensitivity of the rotation to the initial
conditions at this late age.  We interpret this as an indication that
the angular momentum coupling time scale is intermediate between the
Pleiades and Hyades ages, which is consistent with the flat solar
rotation curve.

The recent Orion data set of Stassun \etal (1999) also has interesting
consequences for angular momentum evolution.  They were sensitive to
rotation periods shorter than eight days; this complicates the
question of directly testing the accretion disk-locking model, since
this is near the peak of the period distribution found by earlier
studies of young stars.  However, they found 85 (out of 264) stars
with rotation periods less than 3 days.  By comparison, models with an
initial rotation period of eight days would have a rotation period of
2.64 days at an age of 1 Myr. Stassun \etal (1999) observed a cutoff
in the distribution below 0.5 days, which would correspond to a
rotation period of 2 days at the birthline (a factor of two greater
than the maximum angular momentum we assumed for setting the upper
envelope of the distribution). This indicates that both the initial
period and the distribution of accretion disk lifetimes needs to be
taken into account when modeling the rapid rotator distribution.
Because we use the upper envelope of rotation to set the value of
\wcrit, assigning a shorter initial period to the upper envelope of
rotation would imply systematically larger values of \wcrit.  This
would lead to shorter predicted disk lifetimes for the slower rotators
in young open clusters, although not enough to alter the qualitative
conclusions about differentially rotating versus solid body models.

An additional feature of these models is the possible lack of long
disk lifetimes observed for very low mass stars. This is most obvious
in a plot of disk lifetime as a function of T$_{eff}$, shown in
figures 15 and 16. Both solid body and differentially rotating models
show the same trend. In the Pleiades, we would expect to see stars
with \vsini at or near the lower detection limit of 7 km/s below 3500
K. The low mass portion of the data set was chosen based on colour of
the stars and not on any rotational information, and therefore should
be unbiased (Stauffer \etal 1999). However, we see a lack of slow
rotators at the low mass end.

There are two possible explanations.  One scenario is that the lowest
mass stars have systematically shorter accretion disk lifetimes than
higher mass stars.  A second possibility is suggested by a comparison
of figures 10 and 11 with figures 12 and 13; the mass dependence of
the saturation threshold for angular momentum loss has a strong
influence on the predicted rotation rates for a given initial
condition.  With the small Hyades cool star sample, the upper envelope
appears to be flat; we therefore infer relatively short disk coupling
lifetimes for the lowest mass stars.  However, if the upper envelope
were to rise with decreasing mass in the Hyades this would indicate
that it is the angular momentum loss law, not a mass-dependent set of
initial conditions, that was responsible.  We cannot rule this
explanation out due to the relatively small sample of Hyades stars
cooler than 3500 K.  However, the two solutions make different
predictions for the observed rotation rates in older clusters.  If the
trend towards higher mean rotation rates at lower mass in the Hyades
persists below 3500 K, this indicates that the mass dependence of the
angular momentum loss law is the primary cause; if the data can be fit
using the existing angular momentum loss law with consistently shorter
disk lifetimes for lower mass stars it is an indication of a genuine
change in the distribution of initial conditions as a function of
mass.

\section{Discussion}

In this paper, we present models of very low mass stars ($<$ 0.5
\Msunns) which include the effects of rotation. These models have been
made possible by the work of a number of groups on the physics of low
temperature stellar atmospheres, opacities and equations of state. By
including the physics of many molecules in stellar atmosphere
calculations, Allard \& Hauschildt (1995) have created models which
are valid for low mass stars with effective temperatures less than
4000 K. The equation of state of Saumon, Chabrier \& Van Horn (1995)
includes partial dissociation and ionization of hydrogen and helium
caused by both pressure and temperature effects, and is applicable to
both low mass stars and giant planets. Finally, Alexander \& Ferguson
(1994) added atomic line absorption, molecular line absorption and
some grain absorption and scattering to the usual sources of
continuous opacity to produce opacity tables which reach to
temperatures as low as 700 K. Since most previous atmosphere, opacity
and equation of state did not include the effects of molecules and
grains, these three improvements represent a great leap forward in our
ability to model very low mass stars.

\subsection{The Structural Effects of Rotation}

We have investigated the effect of rotation on the structure of low
mass stars. We discussed a number of implications, based on models
which demonstrate the maximum extent of the differences between
rotating and non-rotating models. The most important effect is the
reduction of effective temperature for stars of a given mass. Rapid
rotators are cooler than slow rotators, and so, for stars more massive
than 0.5 \Msunns, any relationship between temperature and mass should
take into account the rotation rate of the star. We have shown that
the structural effects of rotation in very low mass stars (less than
$\sim$ 0.5 \Msunns) are minimal and can be neglected when interpreting
temperatures and luminosities of these stars from observations.  Table
2 gives the polynomial correction between the effective temperature of
a rotating star and that of a non-rotating star, as a function of
rotation rate and stellar mass.

We have shown that the kinetic energy of rotation is not a significant
contribution to the total luminosity of stars between 0.1 and 1.0
\Msunns, and does not change the timescale for evolution on the
pre-main sequence.  This rotational contribution to the total energy
of the star does not affect the position of the star in the HR
diagram, and we have neglected this effect in the evolutionary
calculations presented in this paper.

Stellar activity can also influence the colour-temperature
relationship; because of the well-known correlation between increased
rotation, increased stellar spot coverage, and increased chromospheric
activity this will tend to change the observed position of rapidly
rotating stars in the HR diagram relative to slow rotators.  Different
colour indices are affected to different degrees; for example, Fekel,
Moffett, \& Henry (1986) found a systematic departure between the B-V
and V-I colours of active stars.  Stars with more modest activity
levels have a more normal colour-colour relationship (e.g.  Rucinski
1987).  Active stars tend to be bluer in B-V than in V-I relative to
less active stars.  Fekel, Moffett, \& Henry (1986) treated this as an
infrared excess, but in open clusters such as the Pleiades and
$\alpha$ Persei rapid rotators are on or above the main sequence in
V-I, but can be below the zero-age main sequence in B-V (Pinsonneault
\etal 1998). Given the theoretical trends presented here, this
suggests that V-I is a good tracer of temperature, and that B-V is the
colour which is most affected by activity. The difference between
effective temperatures based on B-V and those based on V-I can reach
200 K in Pleiades stars (Krishnamurthi, Pinsonneault, King, \& Sills
1999).

\subsection{Angular Momentum Evolution}

The study of low mass stars can provide valuable constraints on the
three coupled ingredients of angular momentum evolution models:
internal angular momentum transport, angular momentum loss, and the
distribution of initial conditions.

We have compared the properties of solid body models with those of
models with internal angular momentum transport from hydrodynamic
mechanisms.  We confirm previous results that the angular momentum
evolution of systems at and younger than the Pleiades age of 110 Myr
are best reproduced by models which permit differential rotation with
depth.  We also find that the solid-body models do a better job of
reproducing the data at the Hyades age (600 Myr) and older; in
addition, helioseismic data are inconsistent with the strong
differential rotation with depth predicted by models with hydrodynamic
angular momentum transport.  The simplest solution that is consistent
with the data is an additional angular momentum transport mechanism
with a time scale intermediate between 110 Myr and 600 Myr.

Angular momentum loss has a strong impact on the rotational history of
low mass stars; the greatest challenge in understanding the angular
momentum evolution of low mass stars has been distinguishing between
the effects of angular momentum loss, internal angular momentum
transport, and the initial conditions.  The combination of deep
convective envelopes and mild angular momentum loss in stars below 0.6
\Msun makes their behaviour insensitive to the treatment of internal
angular momentum transport.  We believe that these stars provide a
simpler laboratory for the study of the two major other ingredients,
namely the loss law and the initial conditions.

Previous work established that an angular momentum loss law which
saturates at a mass-dependent critical value of the stellar angular
velocity is required to reproduce the fastest rotators in young
clusters. We find that the prescription of Krishnamurthi \etal (1997),
in which $\omega_{crit}$ is inversely proportional to the convective
turnover time of the star at 200 Myr (the Rossby scaling), yields a
consistent solution from 0.6 to 1.1 \Msun.  A Rossby scaling
underestimates the mass dependence when extended to the lowest mass
stars, in the sense that the efficiency of angular momentum loss drops
more rapidly than predicted by a Rossby scaling.  Models for stars
below 3500 K ($\sim$ 0.4 \Msunns) do not yield consistent initial
conditions compared to more massive stars even when the normalization
appropriate for the observed upper envelope of the Hyades data is
used.  It is possible that this could reflect a change in the
distribution of initial conditions, but more data are required to rule
out the angular momentum loss rate as a cause.  It is also possible
that our estimate of the global convective overturn time is inaccurate
for the lowest mass stars.  However, it is clear that angular momentum
loss must be occurring, since models which experience no angular
momentum loss are rotating at velocities which are at least a factor
of 2 too large to agree with any of the observations. In general, we
find no evidence for a significant change in the form of angular
momentum evolution and loss when stars become fully convective; all of
the trends observed are smooth as a function of mass without a sharp
break at the fully convective boundary.

This last point has implications for cataclysmic variable (CV)
research.  The orbital periods of these mass-accreting white dwarfs
are observed to have a gap between 2 and 3 hours, in which very few
systems are seen. The accepted interpretation for this gap has been a
sharp reduction in angular momentum loss rate as the stars become
fully convective (see Patterson 1984 for a complete introduction to
CVs, and McDermott \& Taam 1989 for one particular CV model). For
masses higher than $\sim$ 0.3 \Msunns, angular momentum loss through a
magnetic wind (magnetic braking) is assumed to occur in the
mass-losing secondary of the CV system. The angular momentum is lost
from the system entirely, and the secondary continues to fill its
Roche lobe, transferring mass onto the white dwarf. This mass transfer
keeps the secondary out of thermal equilibrium, so the star has a
slightly larger radius for its mass than an isolated star.  When the
star becomes fully convective, however, it is proposed that the
secondary's magnetic field is no longer anchored to the radiative core
of the star, and abruptly ceases to exist. Angular momentum (and hence
mass) is no longer transferred from the secondary, and the secondary is
allowed to contract to its normal main sequence radius. Only when the
system shrinks further, due to gravitational radiation, does the mass
transfer restart, and the system again becomes a CV. We have not seen
any sharp break in the angular momentum loss rates as we move to lower
masses, suggesting that any theory for the period gap of cataclysmic
variables which relies on the cessation of angular momentum loss
requires a mechanism other than the standard magnetic braking.

There are several promising directions for future theoretical studies.
First, there is an increasing database of rotational periods and
velocities in systems with a range of ages and metal abundances.  A
combined analysis of the information in protostars, young clusters,
and intermediate-aged systems will provide interesting constraints on
theoretical models.  The next generation of theoretical models should
incorporate multiple physical mechanisms for internal angular momentum
transport while also relying on the complementary information on
surface mixing.  More sophisticated models of stellar winds and
angular momentum loss will also be needed to investigate the physical
implications of the empirical trends deduced in theoretical studies of
the type we have performed.

We would like to conclude with a plea to observers of young open
clusters.  The observational database for rotation rates of very low
mass stars is quite sparsely populated. We have observations of
stellar rotation rates down to about 0.2 \Msun in the Pleiades and
Hyades, but the data in $\alpha$ Persei, IC 2602 and IC 2391 have only
spotty coverage below $\sim$ 0.6 \Msunns. These young stars provide
constraints on the early spindown of low mass stars. The Hyades is the
oldest cluster in this sample, and many of the different scenarios are
best distinguished from each other at later ages. It would be very
beneficial to this field to determine rotation rates for the lower
mass stars in young open clusters.

\acknowledgements This work was supported by NASA grant
NAG5-7150. A. S. wishes to recognize support from the Natural Sciences
and Engineering Research Council of Canada. We would like to
acknowledge use of the Open Cluster Database, as provided by
C.F. Prosser (deceased) and J.R. Stauffer, and which currently may be
accessed at http://cfa-www.harvard.edu/$\sim$stauffer/, or by anonymous ftp
to cfa0.harvard.edu (131.142.10.30), cd /pub/stauffer/clusters/.

\clearpage

\begin{figure}
\plotone{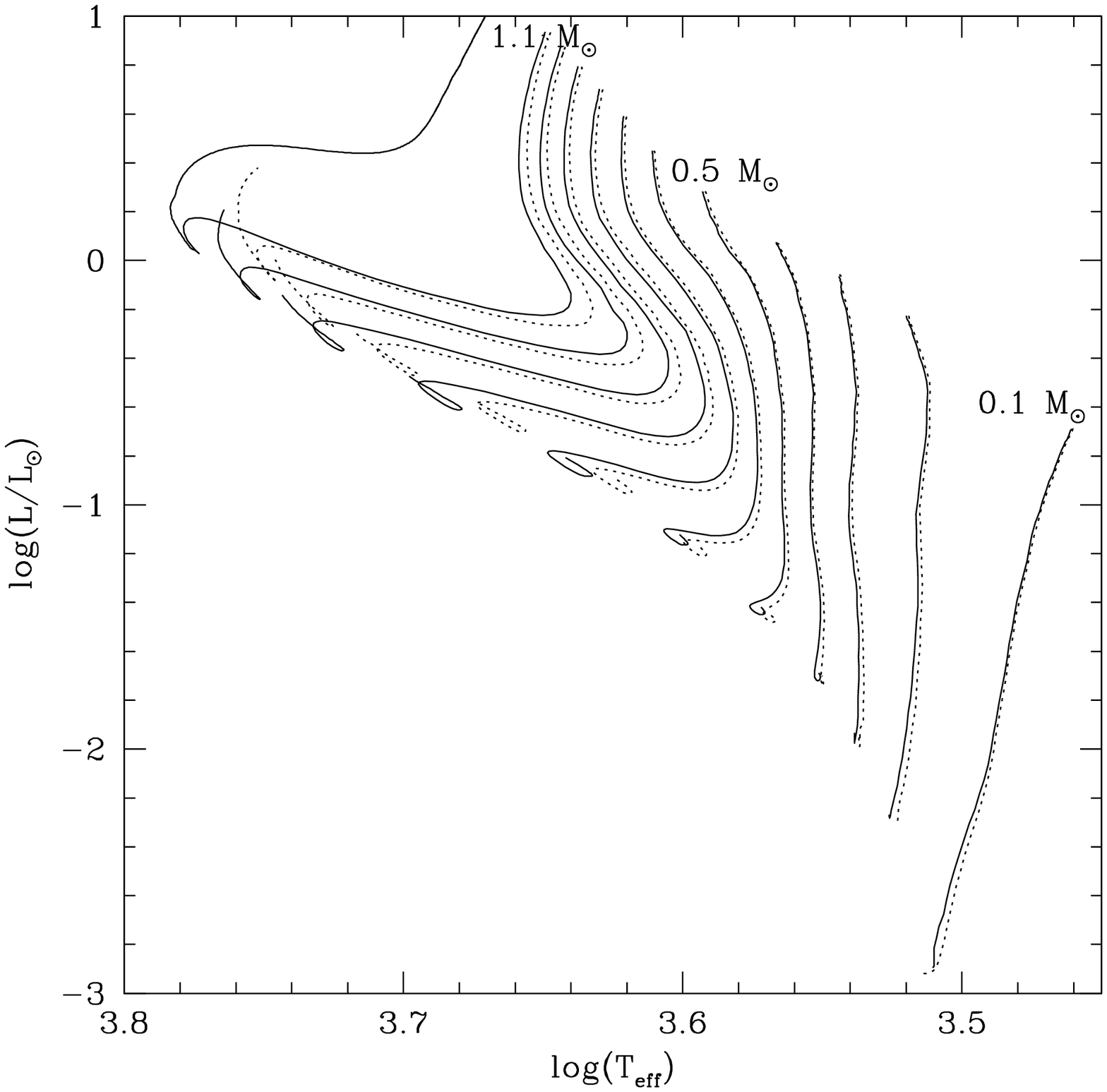}
\caption{Evolutionary tracks for stars with masses
between 0.1 \Msun and 1.1 \Msun in steps of 0.1 \Msunns. The solid lines
are stars without rotation, and the dotted lines are stars which have
initial rotation periods of 8 days, and no angular momentum loss.}
\end{figure}

\begin{figure}
\plotone{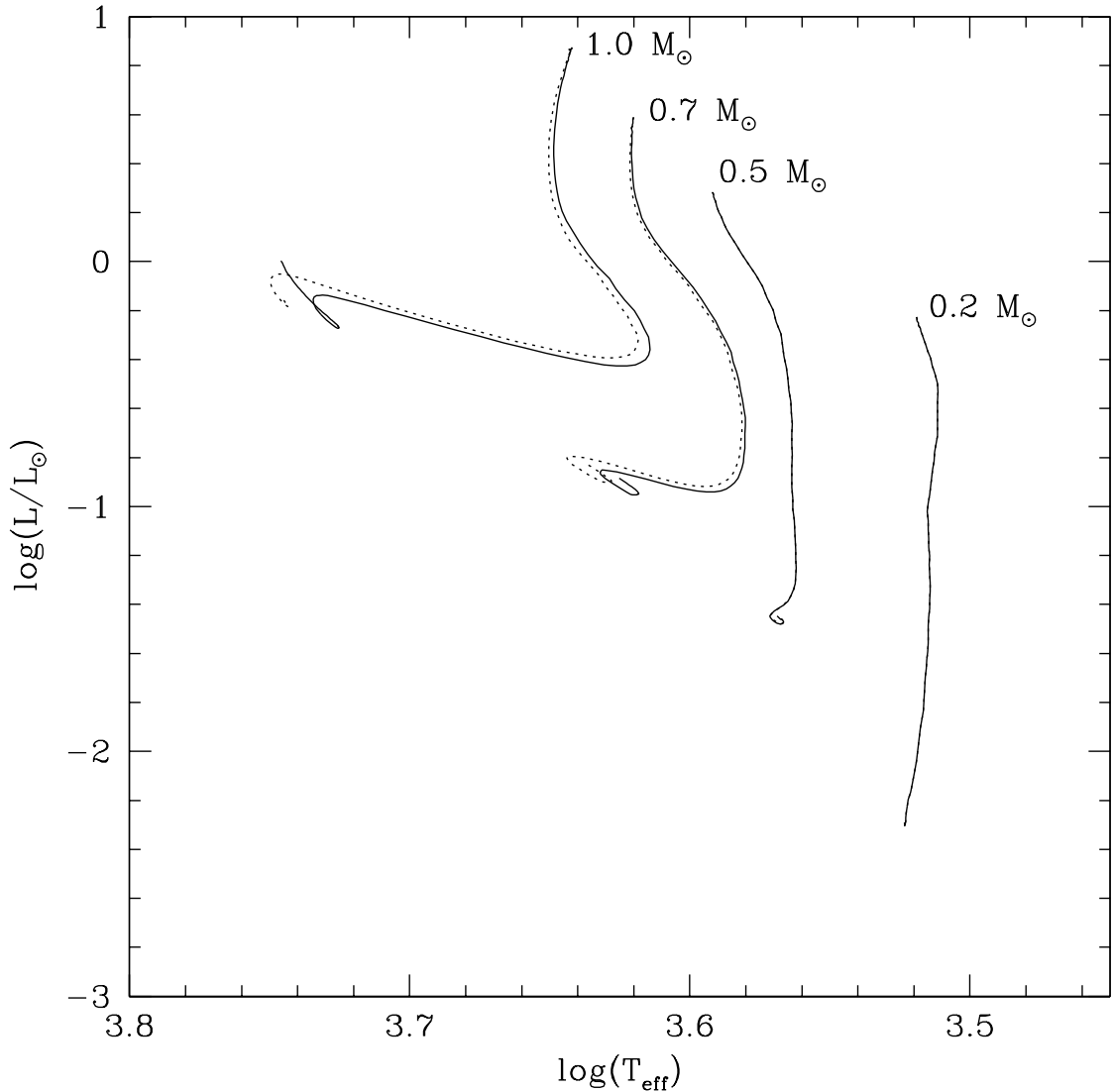}
\caption{Evolutionary tracks for rotating stars
under different assumptions about internal angular momentum
transport. The solid tracks are stars which have differentially
rotating radiative cores and rigidly rotating convection zones, while
the dashed lines show the tracks for stars which are constrained to
rotate as solid bodies. Stars of the same mass have the same surface
rotation rate at the zero age main sequence.  Since the low mass stars
are fully convective throughout their pre-main sequence lifetime, they
always rotate as solid bodies, and so the two tracks are
identical. For the higher mass stars, the tracks are the same while
the star is on the early pre-main sequence. When the star begins to
develop a significant radiative core, however, the two cases have
different evolutionary paths.}
\end{figure}

\begin{figure}
\plotone{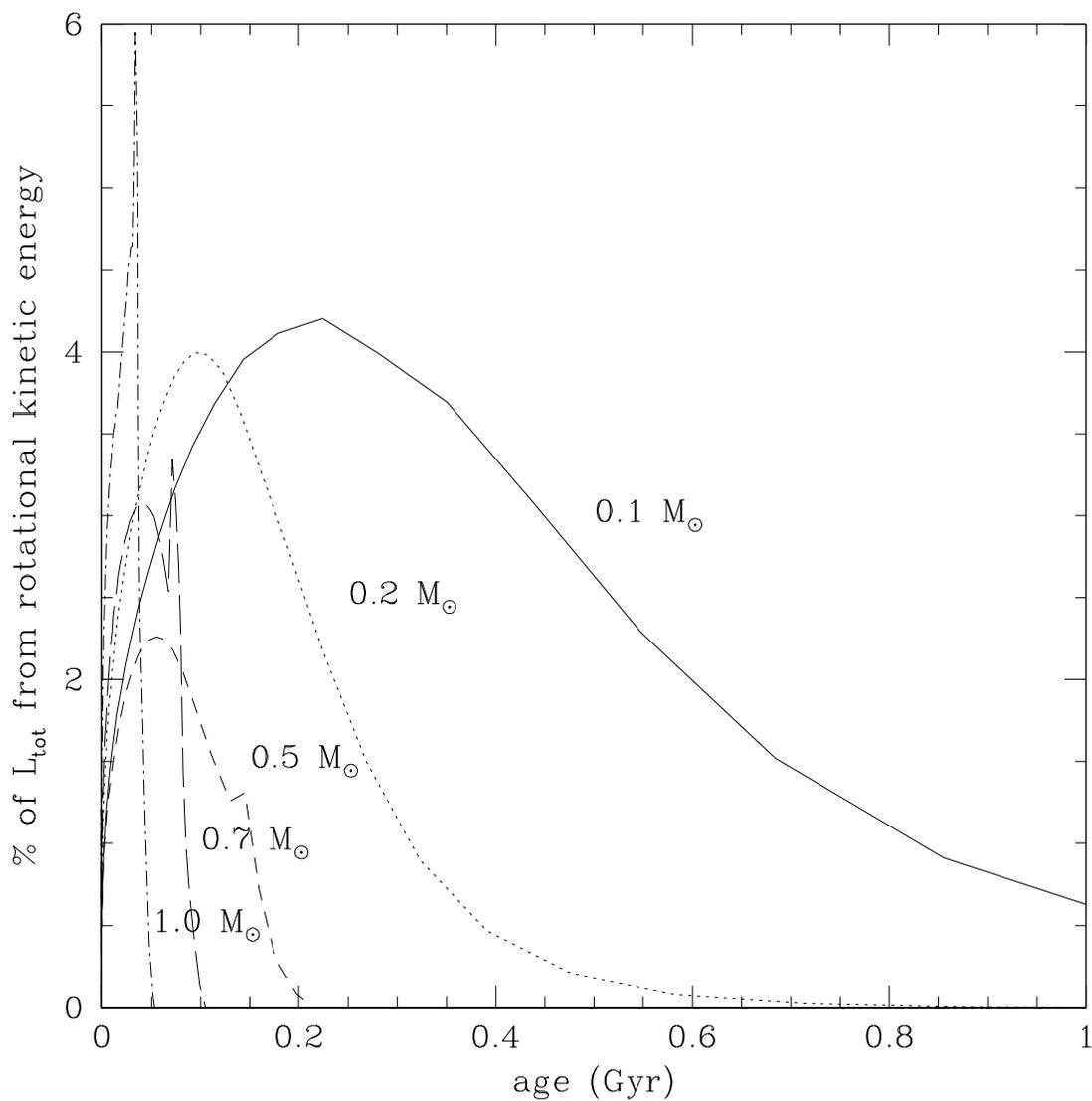}
\caption{The percentage contribution to the total
luminosity by the rotational kinetic energy, as a function of age. The
maximum contribution is 6\%, for the 1.0 \Msun star, but for a short
time. For the 0.1 \Msun star, the contribution of rotation kinetic
energy lasts the longest time. In either case, the effect on the
evolutionary timescale and position in the HR diagram is
insignificant.}
\end{figure}

\begin{figure}
\plotone{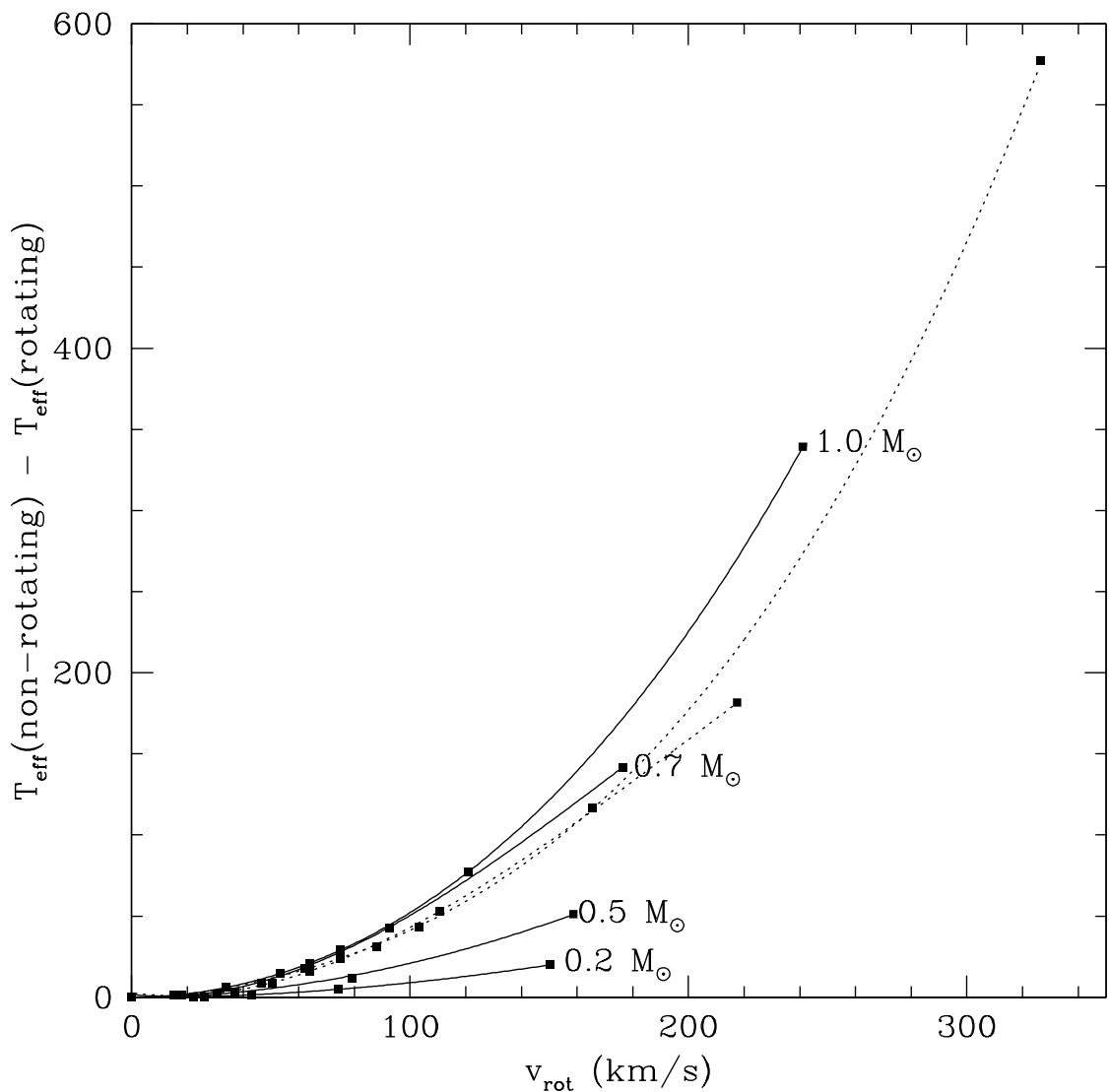}
\caption{Difference between effective temperatures of
rotating and non-rotating stars at the same zero age main sequence, as
a function of surface rotation velocity. The different lines
correspond to stars of different masses, with the highest masses
demonstrating the largest difference in temperature. Models which
allow for differential rotation are plotted using solid lines, while
the solid body rotators are plotted as dashed lines.  A difference of
a 100 K or more will significantly affect the mapping of stellar mass
on observed effective temperature.}
\end{figure}

\begin{figure}
\plotone{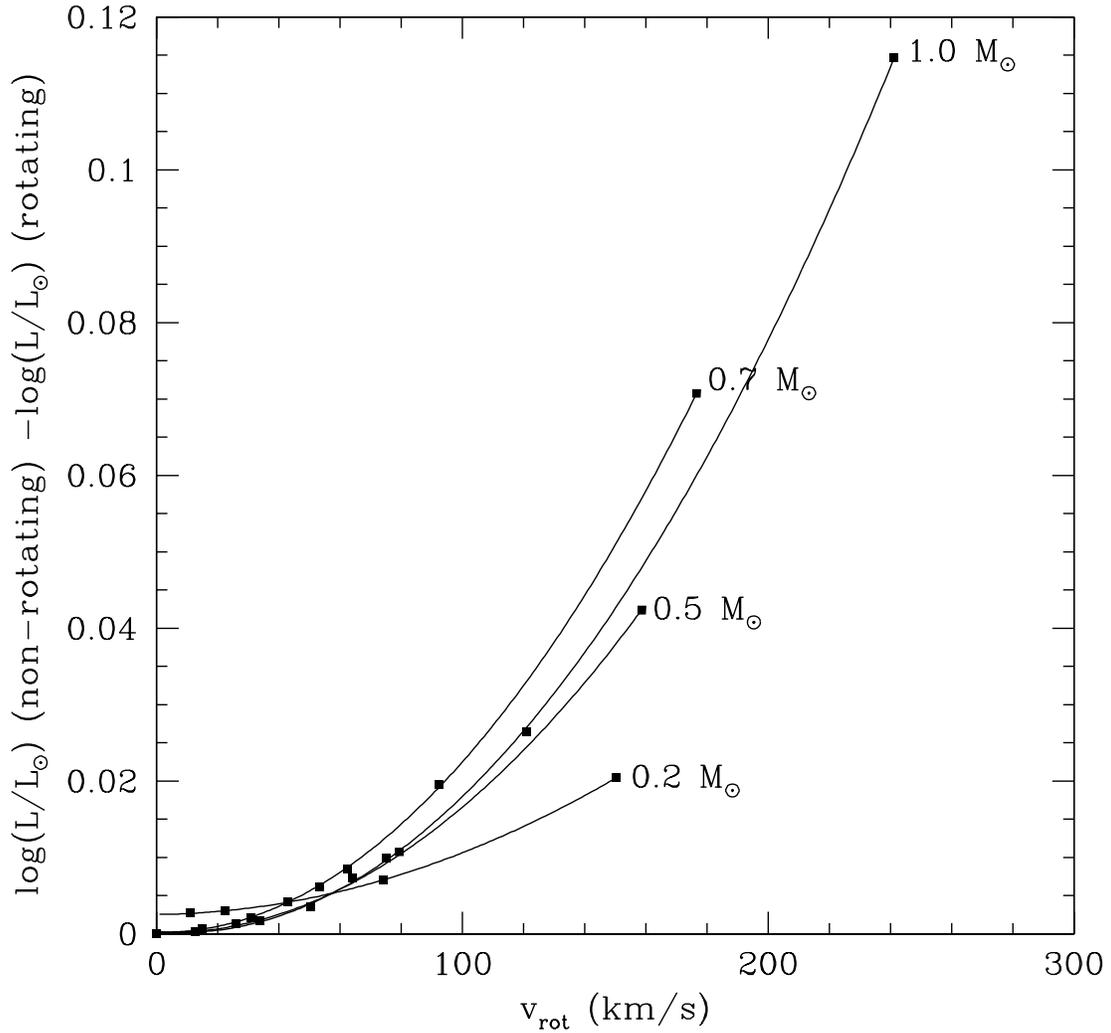}
\caption{Difference between luminosity of rotating and
non-rotating stars at the zero age main sequence, as a function of
surface rotation velocity. The highest mass stars demonstrate the
largest different in luminosity. However, even for 1 \Msunns, the
difference in luminosity is not significant, unlike the difference in
effective temperature caused by rotation.}
\end{figure}

\begin{figure}
\plotone{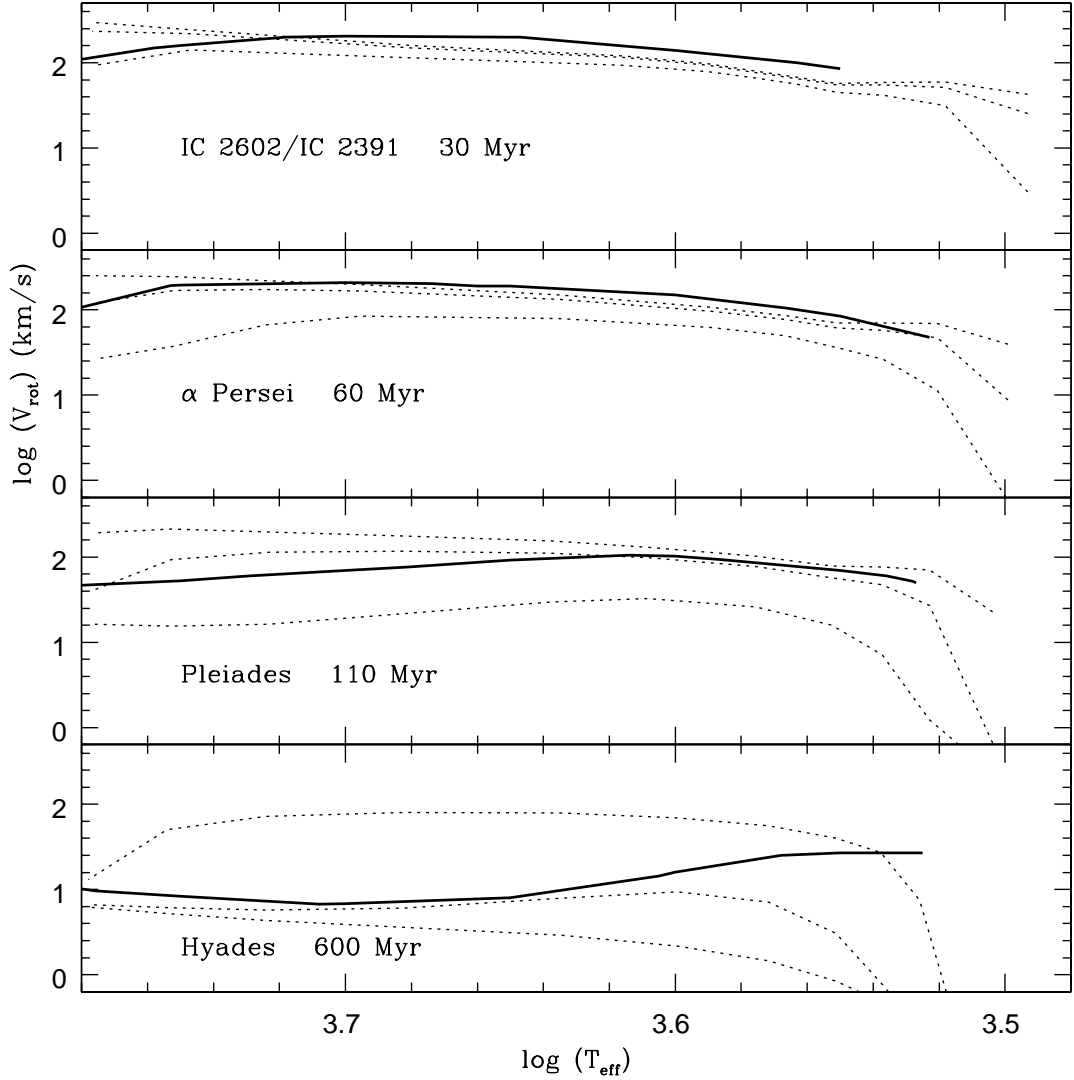}
\caption{Solid body models compared to the upper
envelope of the open cluster data at different ages. A mass dependent
\wcrits was used. Three normalizations are presented here. From top to
bottom of each frame, we used 5, 10 and 20 $\omega_{\odot}$ at 1.0
\Msunns. The thick solid line represents the upper envelope of the
observational data.}
\end{figure}

\begin{figure}
\plotone{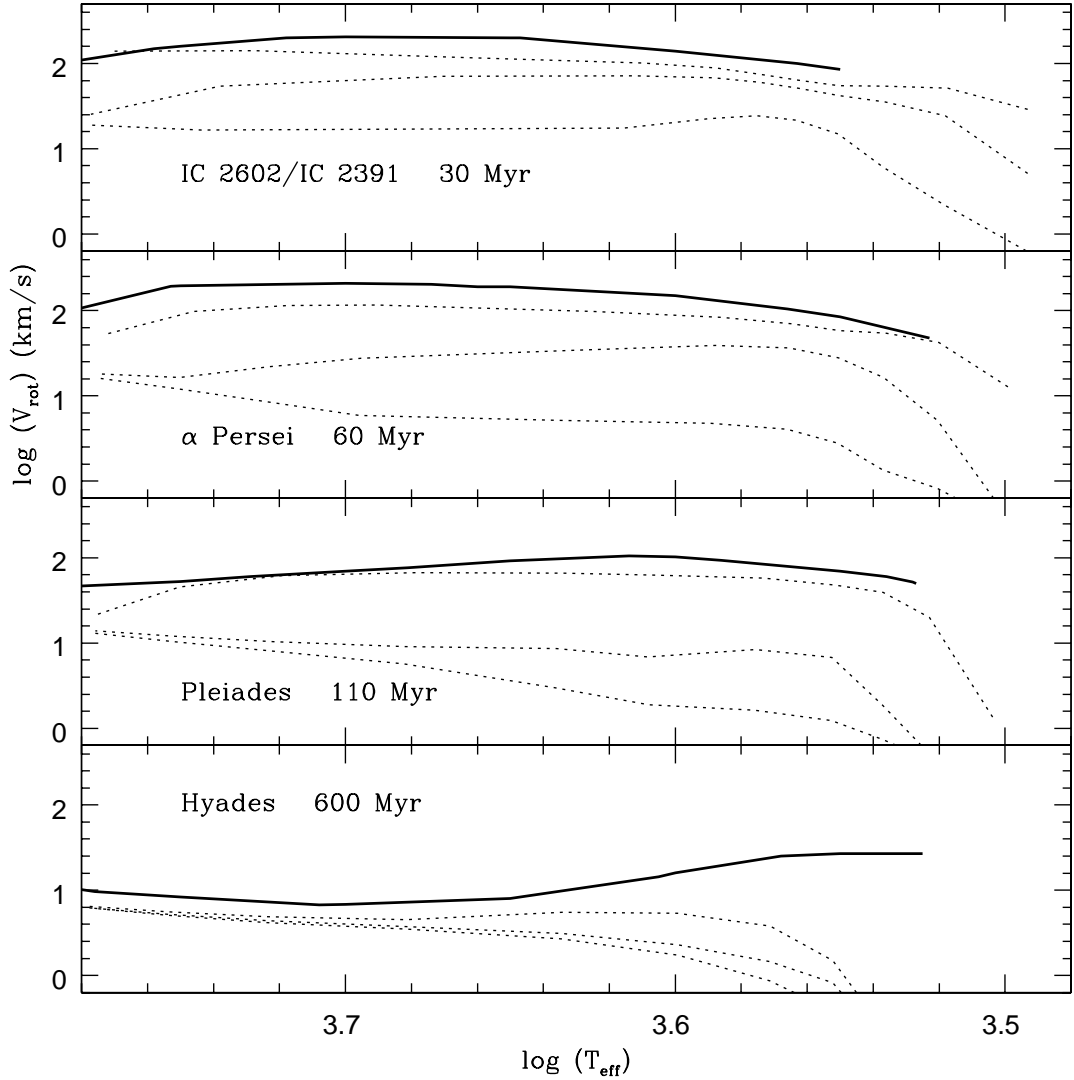}
\caption{As figure 6, with differentially rotating
models. Notice the effect of very rapid rotation in the youngest
clusters on the temperature of the highest mass models. If stars begin
their lives with 4 day periods, the fast rotators in IC 2602 and IC
2391 with temperatures higher than $\log T_{eff}=3.68$ must have
masses higher than 1.1 \Msunns.}
\end{figure}

\begin{figure}
\plotone{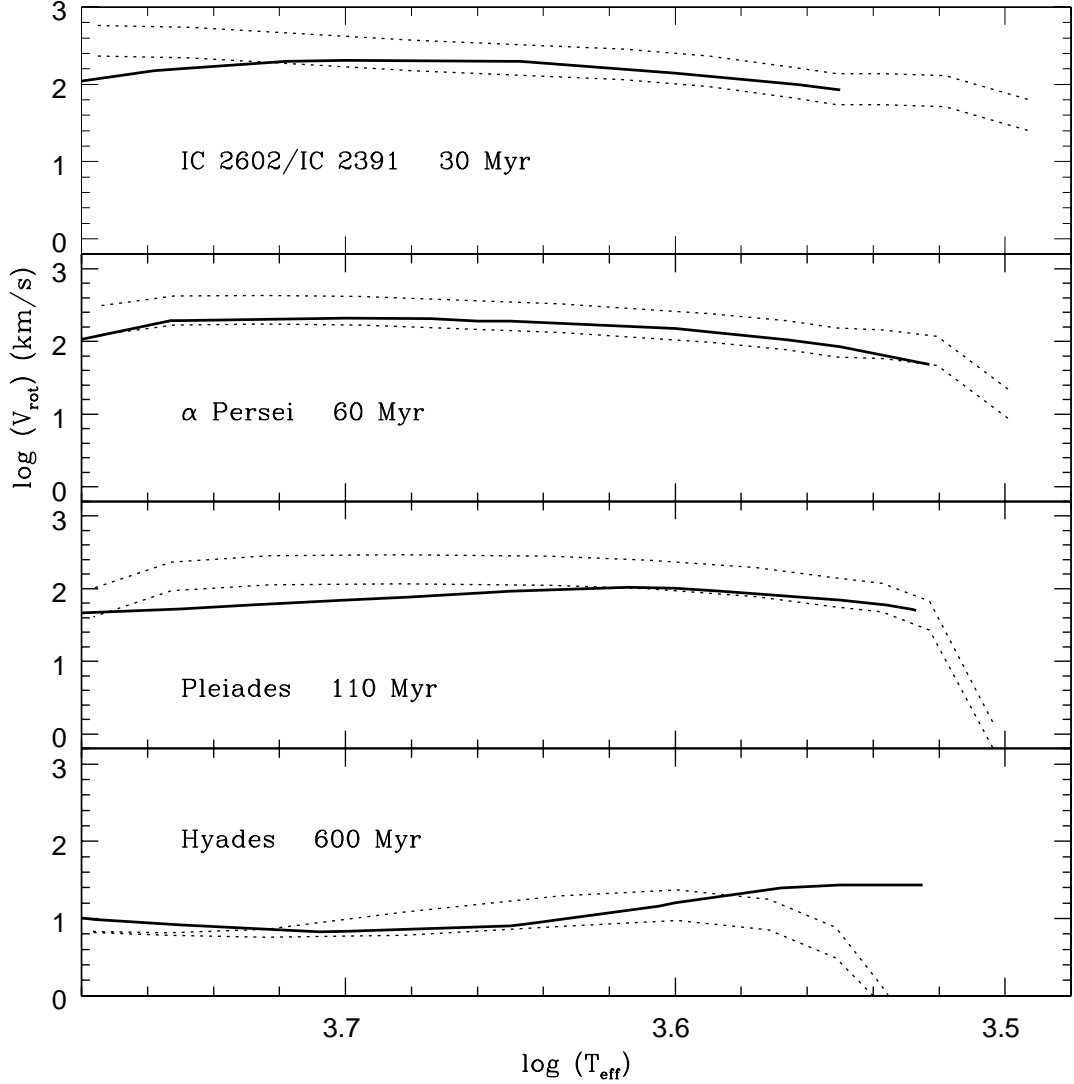}
\caption{Solid body models compared to the upper
envelope of the open cluster data at different ages. A mass dependent
\wcrits was used, normalized to 10 $\omega_{\odot}$ at 1.0
\Msunns. The upper line corresponds to models with an initial rotation
period of 4 days, and the lower line shows models with an initial
rotation period of 10 days.}
\end{figure}

\begin{figure}
\plotone{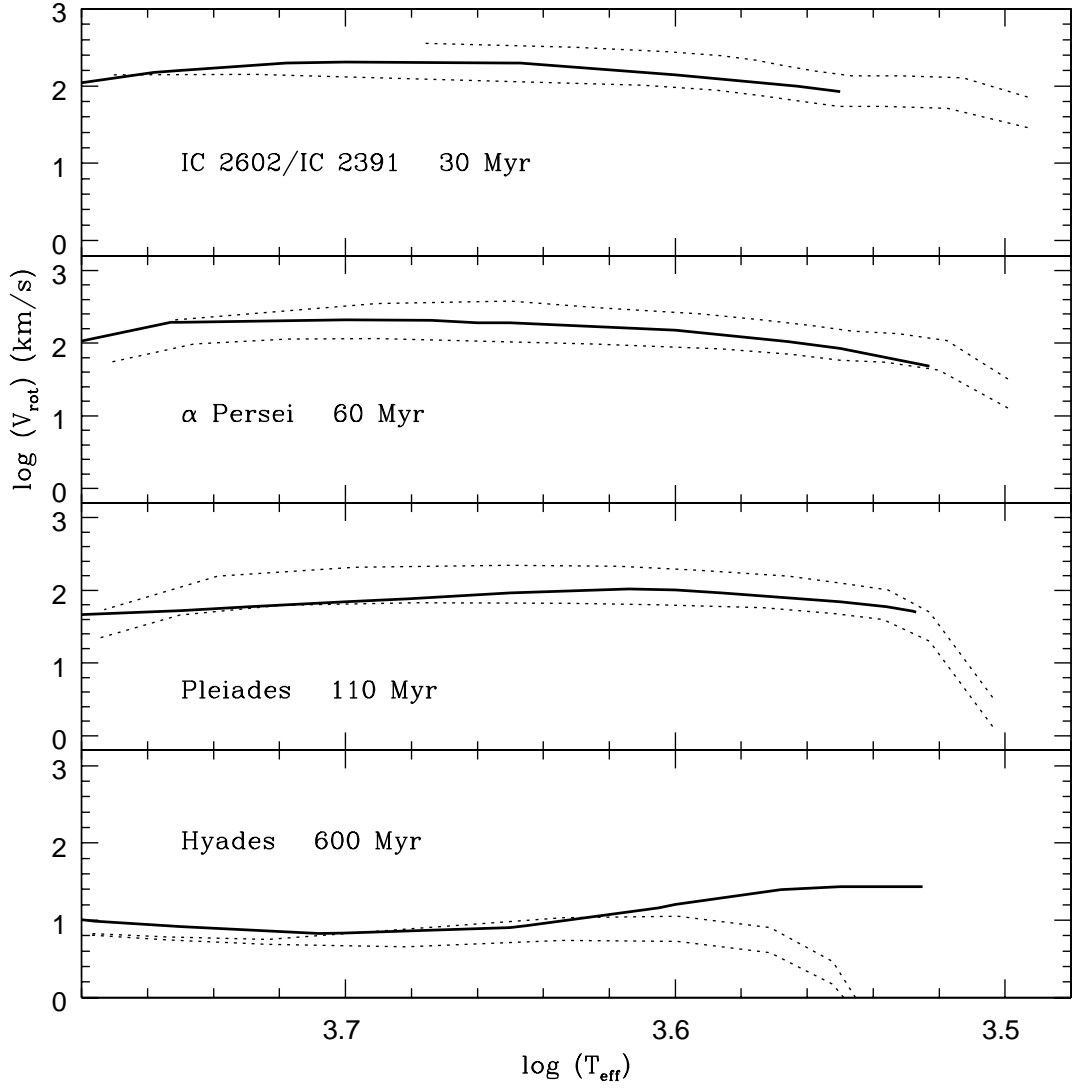}
\caption{As figure 8, with differentially rotating
models. The mass dependent \wcrits was normalized to 5
$\omega_{\odot}$ at 1.0 \Msunns.}
\end{figure}

\begin{figure}
\plotone{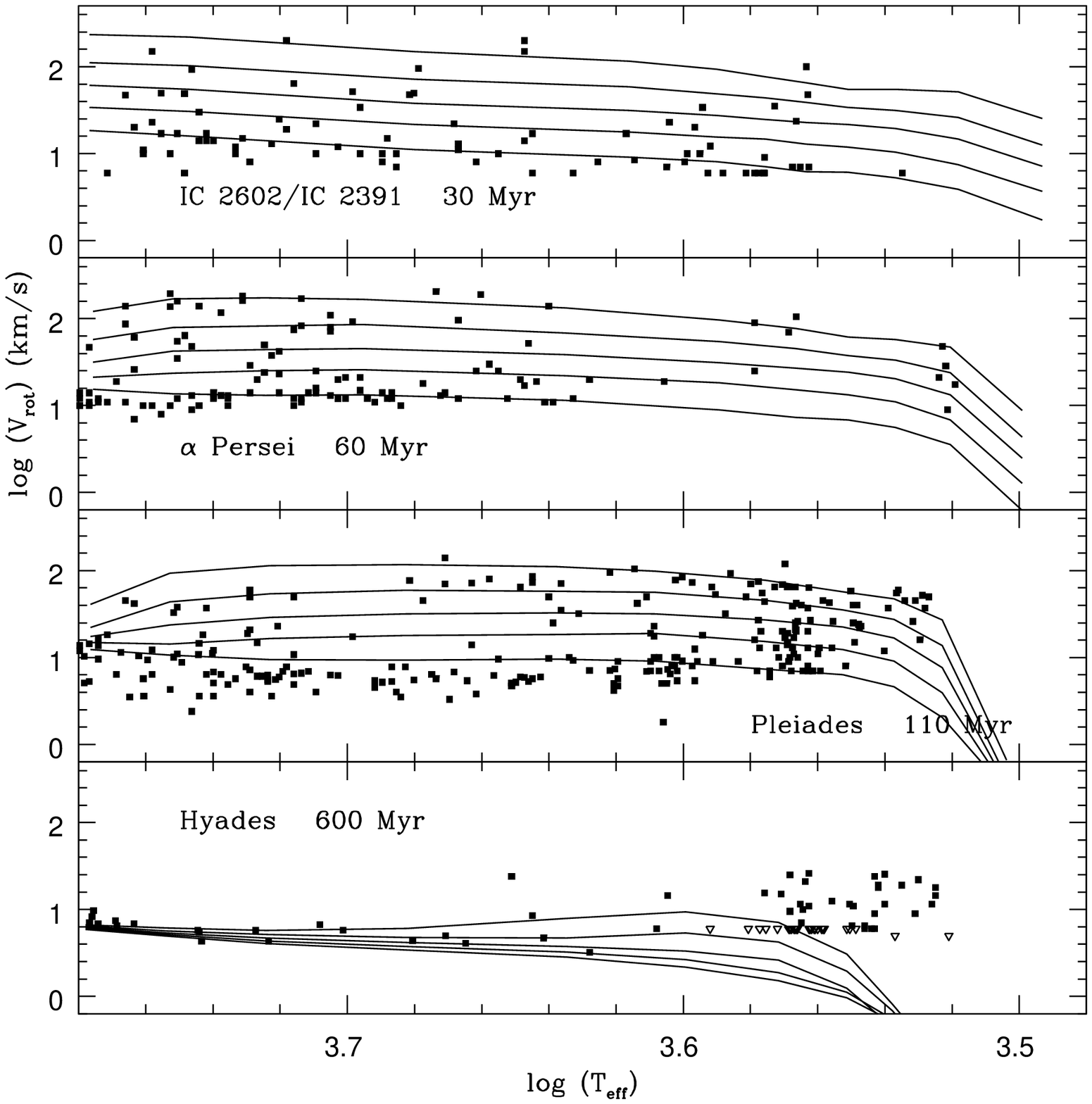}
\caption{Solid body models compared to open cluster data at different ages. A mass dependent
\wcrits was used, normalized to 10 $\omega_{\odot}$ at 1.0
\Msunns. Five different values of the disk locking lifetime were used:
0, 0.3, 1, 3 and 10 Myr (from top to bottom).}
\end{figure}

\begin{figure}
\plotone{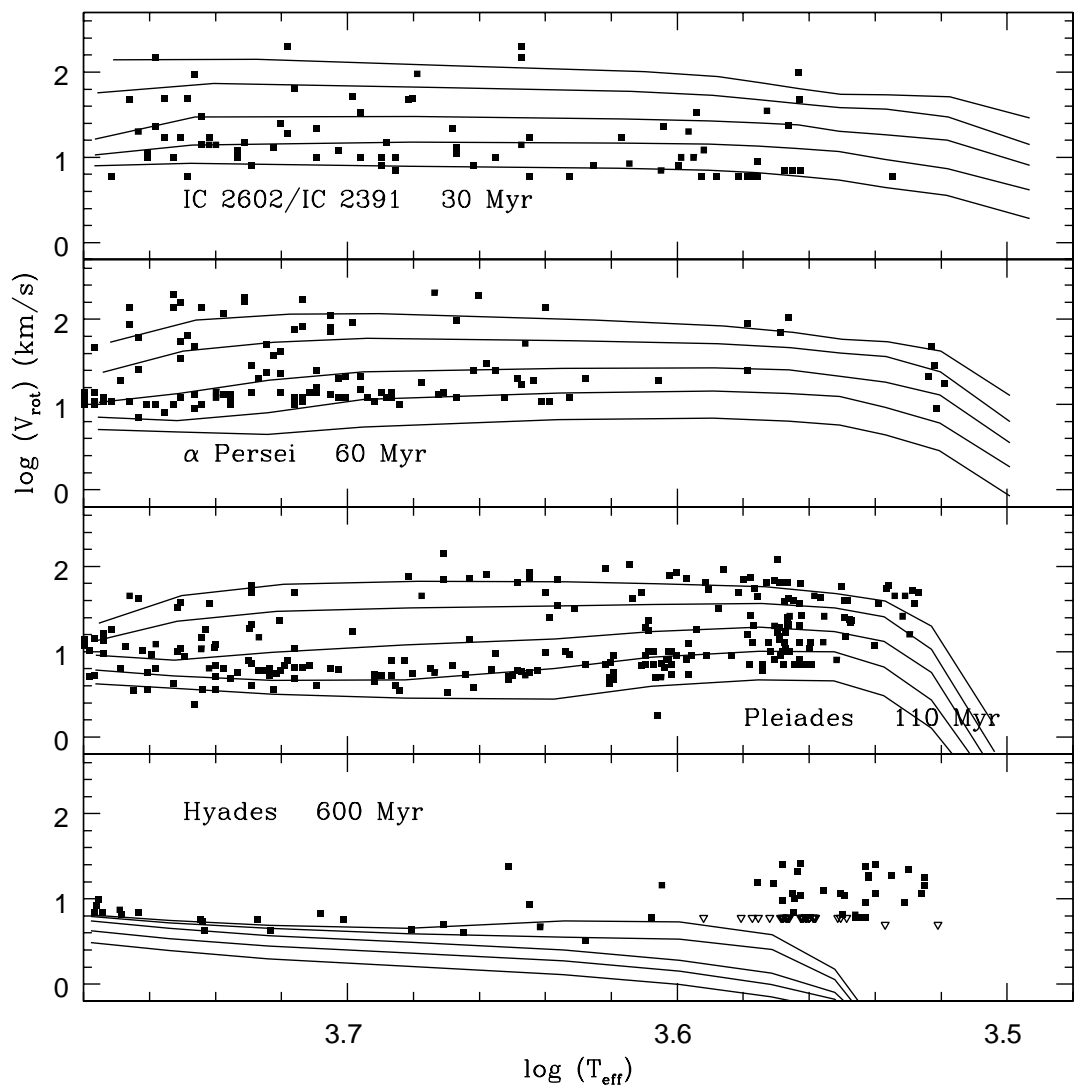}
\caption{Differentially rotating models compared to
 open cluster data at different ages. The
mass dependent \wcrits was normalized to 5 $\omega_{\odot}$ at 1.0
\Msunns. Five different values of the disk locking lifetime were used:
0, 0.3, 1, 3 and 10 Myr (from top to bottom). }
\end{figure}

\begin{figure}
\plotone{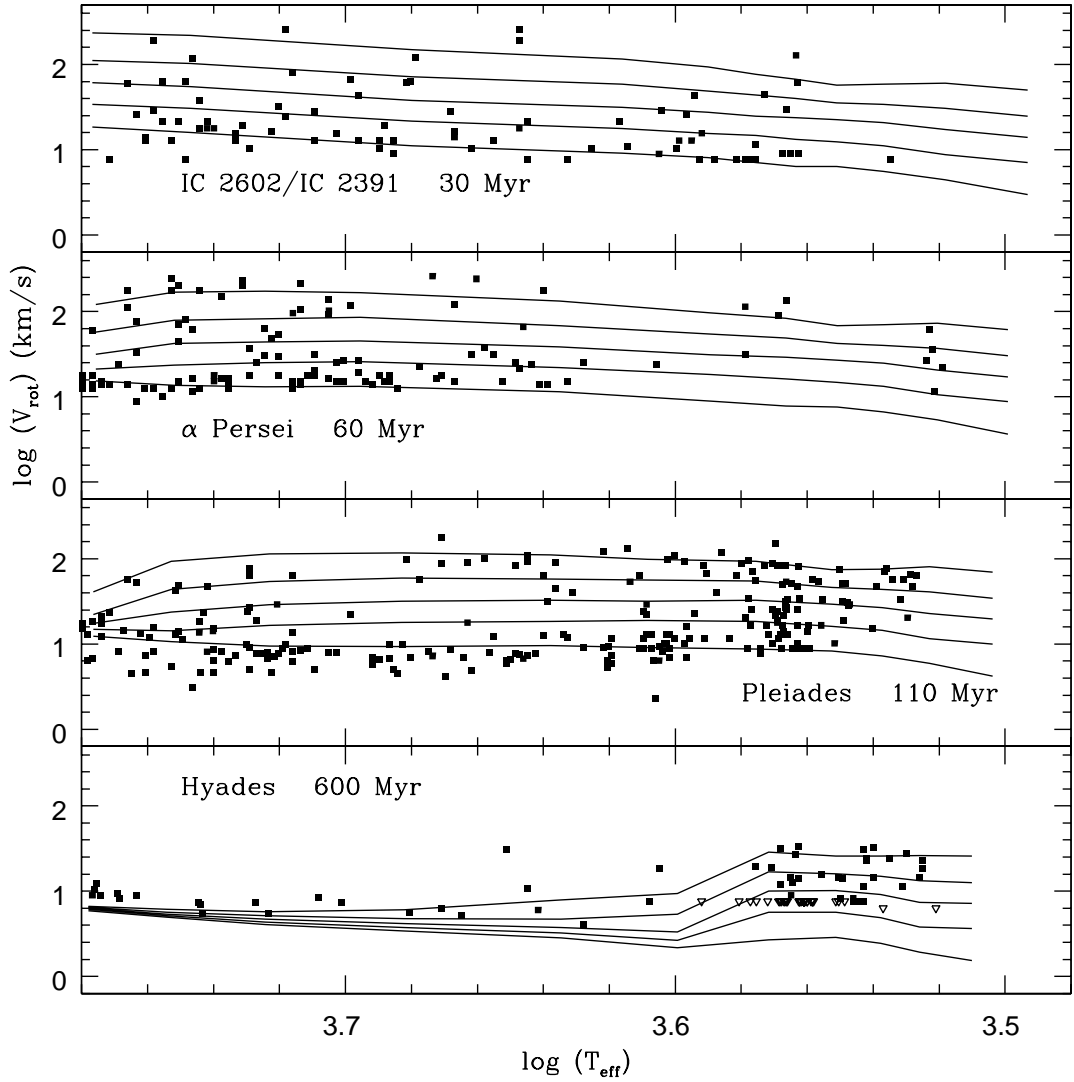}
\caption{Solid body models compared to  open cluster data at different ages.  Five different
values of the disk locking lifetime were used: 0, 0.3, 1, 3 and 10 Myr
(from top to bottom). These models are constrained to more accurately
match the Hyades observations below 0.5 \Msunns.}
\end{figure}

\begin{figure}
\plotone{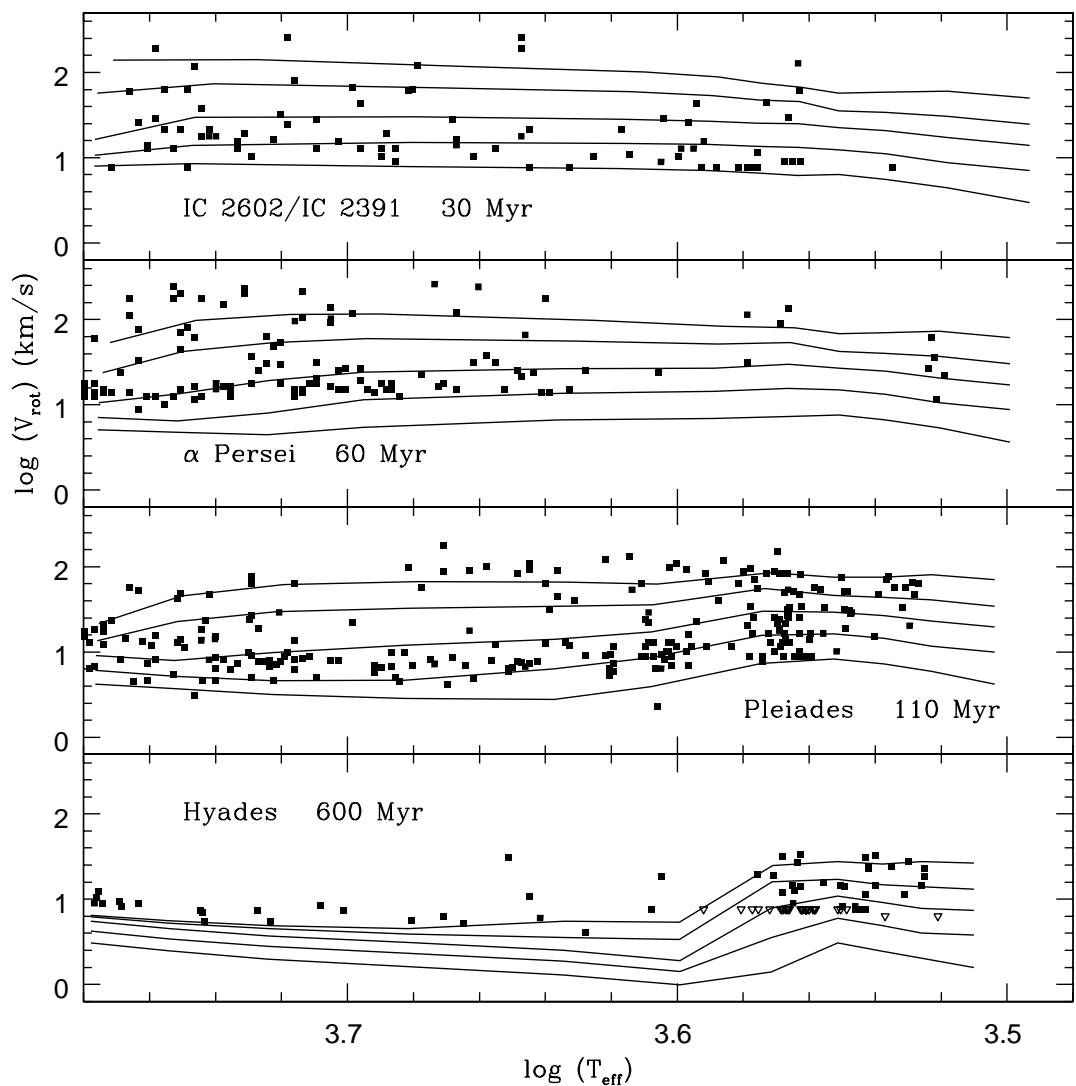}
\caption{Differentially rotating models compared to
open cluster data at different ages.  Five different values
of the disk locking lifetime were used: 0, 0.3, 1, 3 and 10 Myr (from
top to bottom). These models are constrained to more accurately match
the Hyades observations below 0.5 \Msunns.}
\end{figure}

\begin{figure}
\plotone{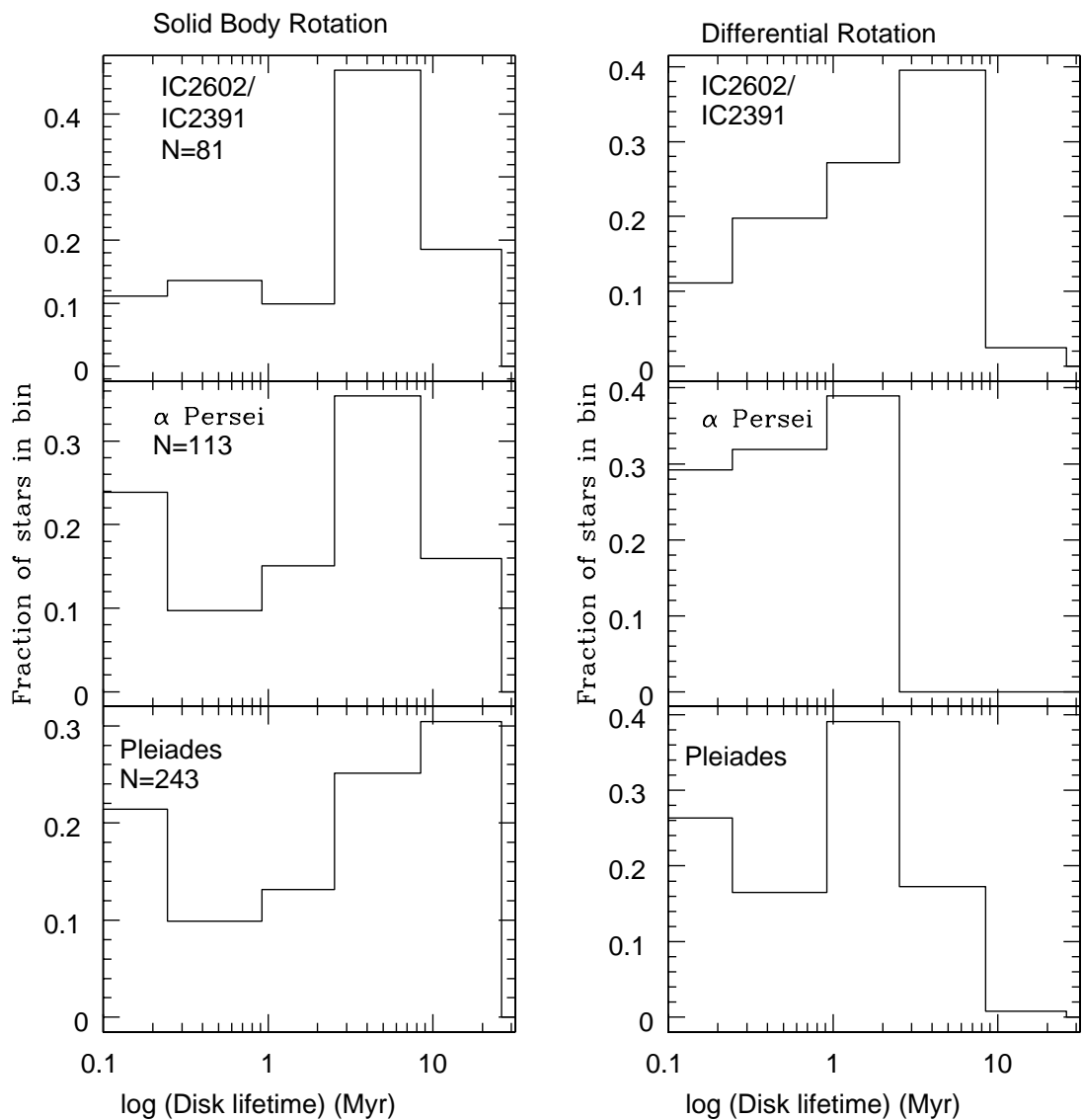}
\caption{Distributions of disk lifetimes for the IC 2602
and IC 2391, $\alpha$ Persei and Pleiades clusters, based on the
models presented in figures 12 and 13. The models which constrain the
stars to rotate as solid bodies require very long disk lifetimes, in
disagreement with observations.}
\end{figure}

\begin{figure}
\plotone{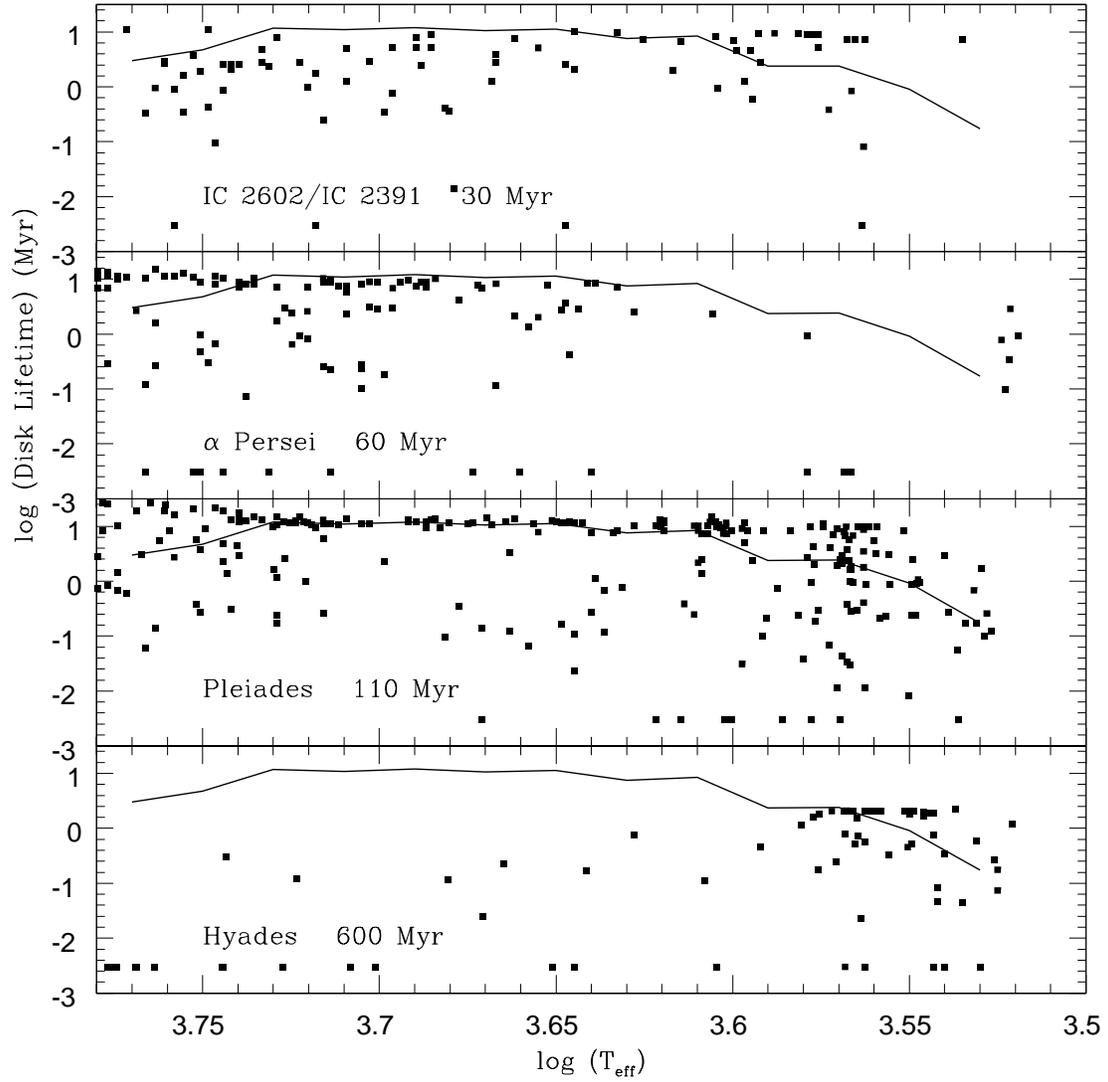}
\caption{Disk lifetimes as a function of effective
temperature, based on the solid body models presented in figure
12. The solid line shows the median of the Pleiades disk lifetimes for
the models which match the Hyades observations (shown
in figure 12). }
\end{figure}

\begin{figure}
\plotone{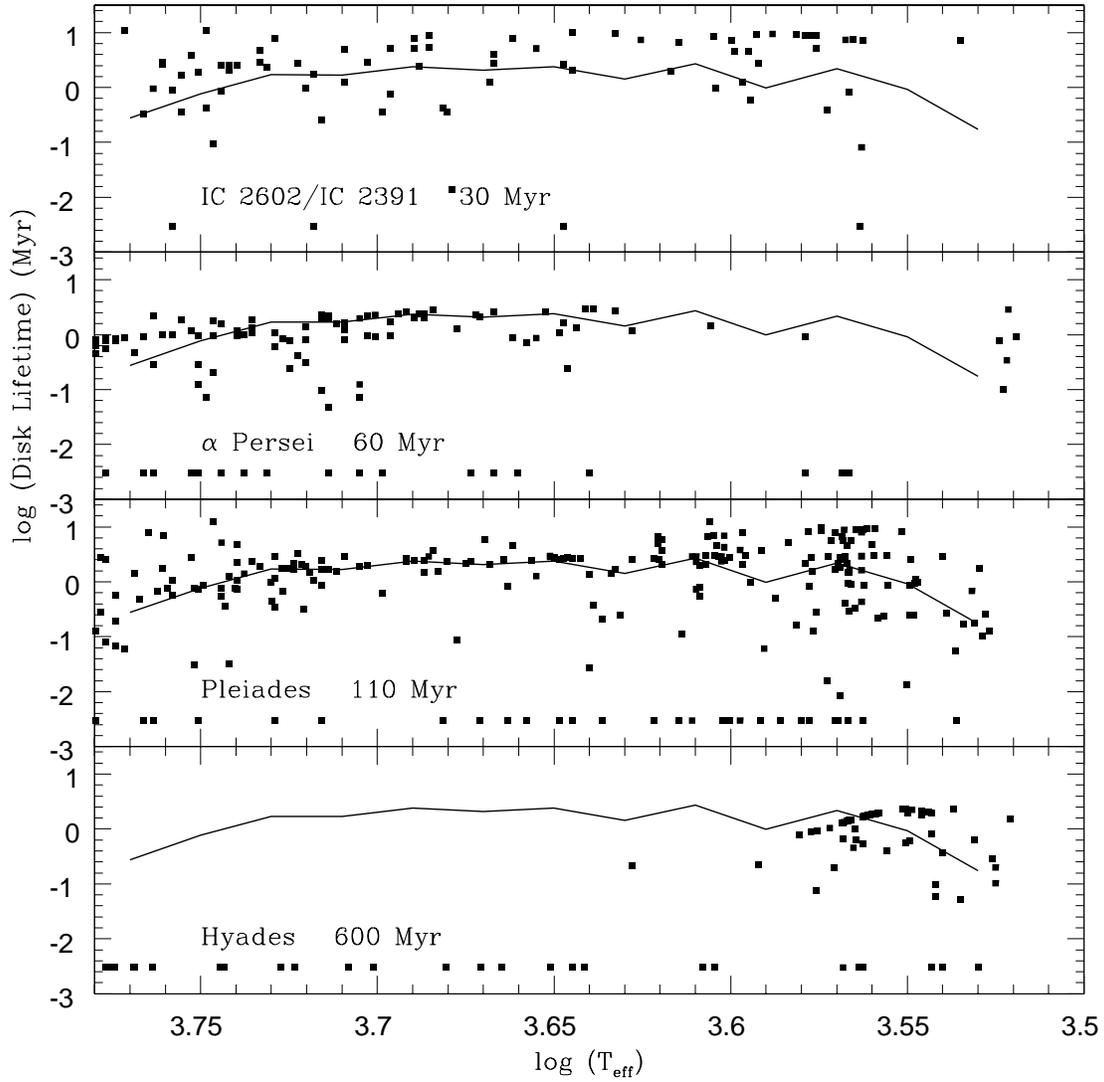}
\caption{As figure 15 with differentially rotating
models. The solid line shows the median Pleiades disk lifetimes as
derived from the model shown in figure 13.}
\end{figure}

\begin{figure}
\plotone{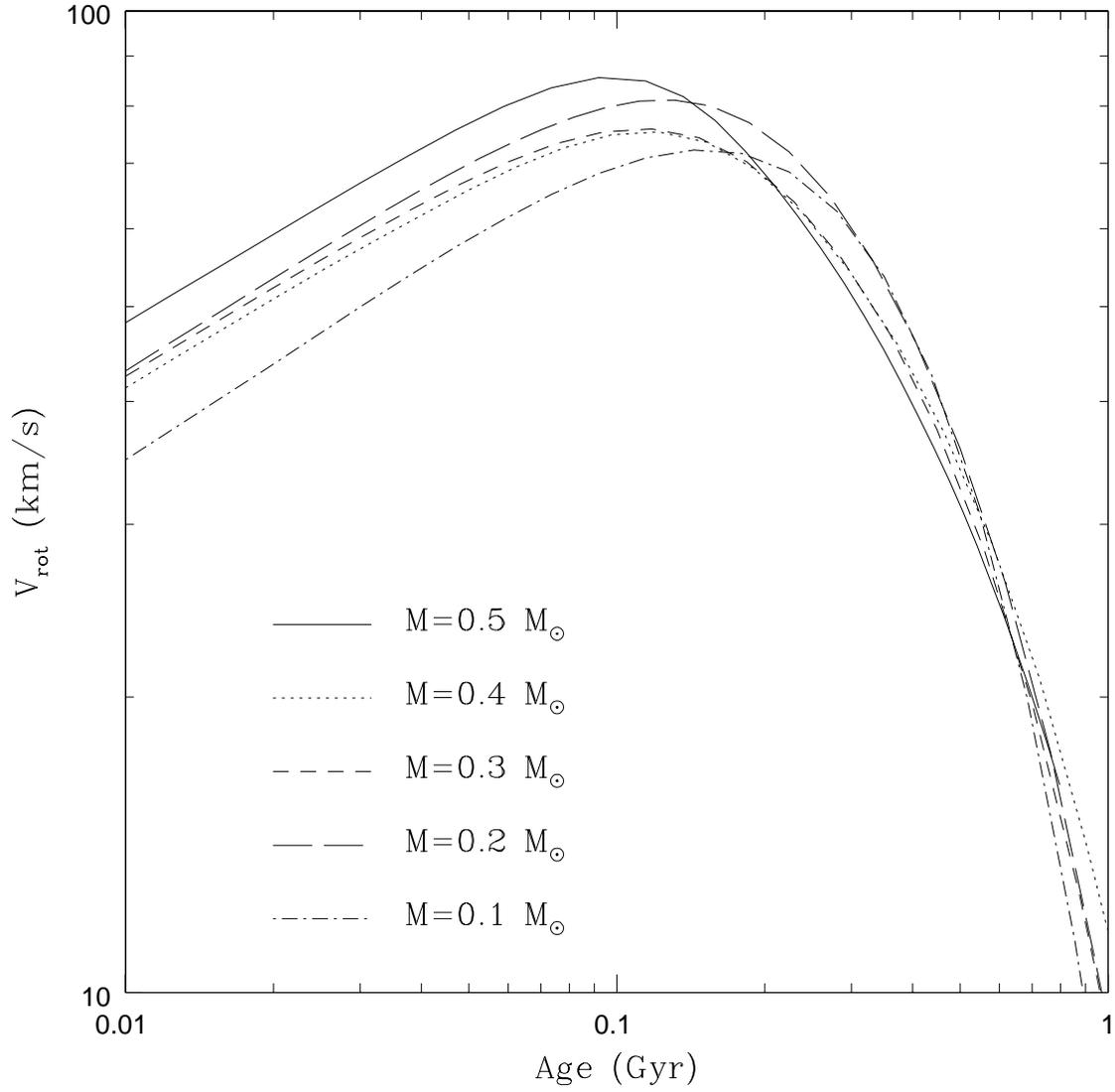}
\caption{Rotation velocity as a function of time for
the low mass stars (M $\leq$ 0.5 \Msunns) from the preferred models
shown in figure 13. The models shown here include no disk locking.}
\end{figure}

\clearpage

\begin{deluxetable}{cccc}
\tablecaption {Zero Age Main Sequence Information for Rotating Models with an
Initial Period of 8 Days, Differential Rotation and no Angular
Momentum Loss}
\tablehead{
\colhead {Mass (\Msunns)} & \colhead {$\log T_{eff}$}  & \colhead
{$\log(L/L_{\odot})$} & \colhead{Age (Myr)} \nl
}
\startdata
0.2 & 3.526 & -2.282 & 890  \nl
0.5 & 3.572 & -1.450 & 280  \nl
0.7 & 3.634 & -0.881 & 230  \nl
1.0 & 3.752 & -0.159 &  27  \nl
\enddata
\end{deluxetable}

\begin{deluxetable}{ccccc}
\tablecaption{Polynomial Coefficients: $T_{eff}^{no rot} -
T_{eff}^{rot} = Av_{rot}^3 + Bv_{rot}^2 + Cv_{rot}+D$ }
\tablehead{
\colhead{Mass (\Msunns)} &  \colhead {A} & \colhead{B} & \colhead{C} &
\colhead{D} 
}
\startdata
0.2 &  $-9.43 \times 10^{-7}$ & $1.08 \times 10^{-3}$ & $-7.71 \times 10^{-3}$ & -0.123 \nl 
0.5 & 0.0 & $1.87 \times 10^{-3}$ & $2.98 \times 10^{-2}$ & -0.746  \nl 
0.7 & $-1.49 \times 10^{-5}$ & $8.13 \times 10^{-3}$ & -0.176 & 1.62 \nl 
1.0 & $ 5.56 \times 10^{-6}$ & $4.37 \times 10^{-3}$ & $3.19 \times 10^{-2}$ &  -0.307 \nl
0.7 solid body & $-1.22 \times 10^{-5}$ & $7.43 \times 10^{-3}$ & -0.217 &  2.63 \nl
1.0 solid body & $ 1.02 \times 10^{-5}$ & $1.53 \times 10^{-3}$ &  0.184 & -2.66 \nl
\enddata
\end{deluxetable}

\end{document}